\documentclass[
reprint, onlinecite, superscriptaddress,
amsmath,amssymb,
aps,prl
]{revtex4-1}

\usepackage{makecell}
\usepackage{graphicx}
\usepackage{dcolumn}
\usepackage{bm}
\usepackage{longtable}
\usepackage{physics}
\usepackage{amsmath}
\usepackage{multirow}

\begin{document}
\title{Intrinsic Spin Photogalvanic Effect in Nonmagnetic Insulator}
\author{Ruixiang Fei}
\affiliation{Department of Physics, Washington
University in St Louis, St Louis, Missouri 63130,
United States}%
\author{Xiaobo Lu}
\affiliation{Department of Physics, Washington
University in St Louis, St Louis, Missouri 63130,
United States}%
\author{Li Yang}
\affiliation{Department of Physics, Washington
University in St Louis, St Louis, Missouri 63130,
United States}
\affiliation{Institute of Materials Science and
Engineering, Washington University in St. Louis,
St. Louis, Missouri 63130, United States }%

\begin{abstract}
We show that with the help of 
spin-orbit coupling, nonlinear light-matter
interactions can efficiently couple with spin
and valley degrees of freedom. This unrevealed
spin photogalvanic effect can drive the long-time pursued
intrinsic pure spin current (PSC) in noncentrosymmetric, nonmagnetic 
insulators. Different from the spin and valley
Hall effect, such a photo-driven
spin current is universal and can be generated without bias. 
Using first-principles simulation, we take 
monolayer transition metal dichalcogenides (TMDs) 
to demonstrate this effect and confirm an enhanced PSC under
linear polarization. The amplitude of PSC is one order larger than the observed charge current in monolayer TMDs. This exotic nonlinear light-spin interaction indicates that
light can be utilized as a rapid fashion to manipulate
spin-polarized current, which is crucial for future 
low-dissipation nanodevice.
\end{abstract}

\maketitle

\textbf{\textit{Introduction.}}-- 
Spintronics is promising for developing the next
generation of energy-efficient electronics because 
it reduces Joule heating and Oersted fields while
retains the functionality of spin currents 
to manipulate magnetization
\cite{sarma04RMP,bader2010spintronics,wolf2001spintronics,Marrows04prl}. 
Pure spin current (PSC), in which 
electrons with different spins travel in 
opposite directions and there is no net motion 
of charge \cite{bhat2005pure}, 
is highly desired for 
spintronic devices\cite{demidov2012magnetic,yang2008giant}.
To date, the generation of PSC in a rapid and
controllable manner represents a challenging 
research direction \cite{awschalom2007challenges}. 
PSC has been induced by the means of spin 
pumping \cite{Gilbert01prl,kajiwara10nature,sandweg11spin,kimel04laser}, 
spin Seebeck\cite{uchida2008seebeck,jaworski2010seebeck,qu13Seebeck}, 
spin Nernst effect\cite{meyer17nernst,sheng17Nernst},
and the spin Hall effect \cite{dyakonov71current,hirsch99spin,kato04observation,wunderlich05exper,huang17bending}. 
Nevertheless, for spin pumping, spin Seebeck or spin Nernst,
it is difficult to realize 
well-localized microwave and terahertz fields 
or temperature gradients at the nanoscale 
in a rapid fashion. Alternatively, the 
spin Hall effect can coverts a longitudinal 
charge current to a transverse PSC 
via extrinsic spin-dependent Mott 
scatterings \cite{dyakonov71current,hirsch99spin} 
or the spin-orbit coupling (SOC) \cite{murakami03hall,sinova04prl,xiao12prl}. 
However, a charge current is required, and 
Joule heating is not fundamentally avoided.

Light-matter interactions is a fundamental 
topic of condensed matter physics. 
Light has been utilized as a powerful tool 
to rapidly and precisely manipulate a wide 
range of properties. Particularly, the spin-polarized 
current accompanied by a net 
charge current\cite{spin01prl,ganichev02spin} 
or the PSC \cite{stevens03quantum,hubner03direct} 
was observed by applying circularly
polarized radiation on quantum interference 
systems, e.g. quantum wells. More recently,
the photo-induced PSC was observed 
in a heterostructure composed of a 
platinum layer and a yttrium-iron garnet
(YIG) film \cite{ellsworth16photo}. 
However, all the above spin currents 
were generated in complex quantum 
systems and require extrinsic mechanisms, 
such as interface engineering and the 
proximity effect. For example, the 
photo-spin current in platinum
is originated from 
the magnetic insulator YIG \cite{ellsworth16photo},
which is not an intrinsic effect of a single crystal.
Therefore, finding a fundamentally new and intrinsic 
mechanism of light-spin interactions is essential
to overcome these challenges. 

Second-order nonlinear optical (NLO) 
responses have been known to be able 
to create DC current or photovoltage in
noncentrosymmetric semiconductors. For example,
shift current is the microscopic mechanism 
of the linear bulk photovoltaic (BPV) 
effect\cite{von81prb,young12first,sipe00second},
while injection current is the origin
of the circular photogalvanic 
effect \cite{sipe00second,morimoto16prb,de2017quantized} 
in time-reversal invariant materials. 
Nevertheless, those studies have focused on
charge currents.How light interacts with spins of electrons is mostly unknown, and corresponding theoretical framework have yet established.
In this work, we find that spins and 
valleys of electrons can strongly couple
with light via second-order optical 
responses in noncentrosymmetric materials. 
This mechanism can selectively generate 
PSC or spin polarized current in nonmagnetic 
materials, which we called it the
“intrinsic spin photogalvanic effect”.

\textbf{\textit{Theory and Mechanism.}}-- 
We consider time-reversal invariant noncentrosymmetric
semiconductors with sizable SOC. We use monolayer
transition metal dichalcogenides (TMDs) 
to demonstrate our theory. The typical band structures
around band edges are shown in Figs. 1a and 1b. 
The spin-up and spin-down bands at $K$/$-K$ points
of reciprocal space are split by SOC and the 
spin order is reversed due to time-reversal 
symmetry. Under coherent light illumination,
the second-order nonlinear DC 
photocurrent density is
\begin{align}
\begin{split}
\label{eq1}
J_{c}=\chi_{a b c}(0 ; \omega,-\omega) E_{a}(\omega) E_{b}(-\omega)
\end{split}
\end{align}

where $E(\omega)$ is the electrical field 
of incident light with a frequency $\omega$, 
indices $a$ and $b$ are the light polarization
directions, and c is the current direction.
The $\chi_{abc}(0;\omega,-\omega)$
is the second-order current susceptibility, 
which has contributions from interband
excitation, so-called
shift current and injection current, 
and intraband nonlinear process, namely 
the optical rectification or the nonlinear
Hall conductivity \cite{sodemann15quantum,ma19Nhall,kang19nonlinear,Shao20prl,wang19Hall,singh20arxiv}.
For semiconductors, the intraband nonlinear process
can be neglected because of the Berry-curvature 
dipole ($\partial_{k} \Omega^{c}$)
is zero at the chemical potential located 
within the bandgap. Thus, we focus on the 
leading-order interband contributions in the
following, i.e., the shift-current mechanism
($\chi_{abc}(0;\omega,-\omega)\equiv\sigma_{abc}(0; \omega,-\omega))$
and injection-current mechanism 
($\chi_{abc}(0;\omega,-\omega)\equiv\eta_{abc}(0; \omega,-\omega)$). 

\begin{figure}
\includegraphics[width=9cm]{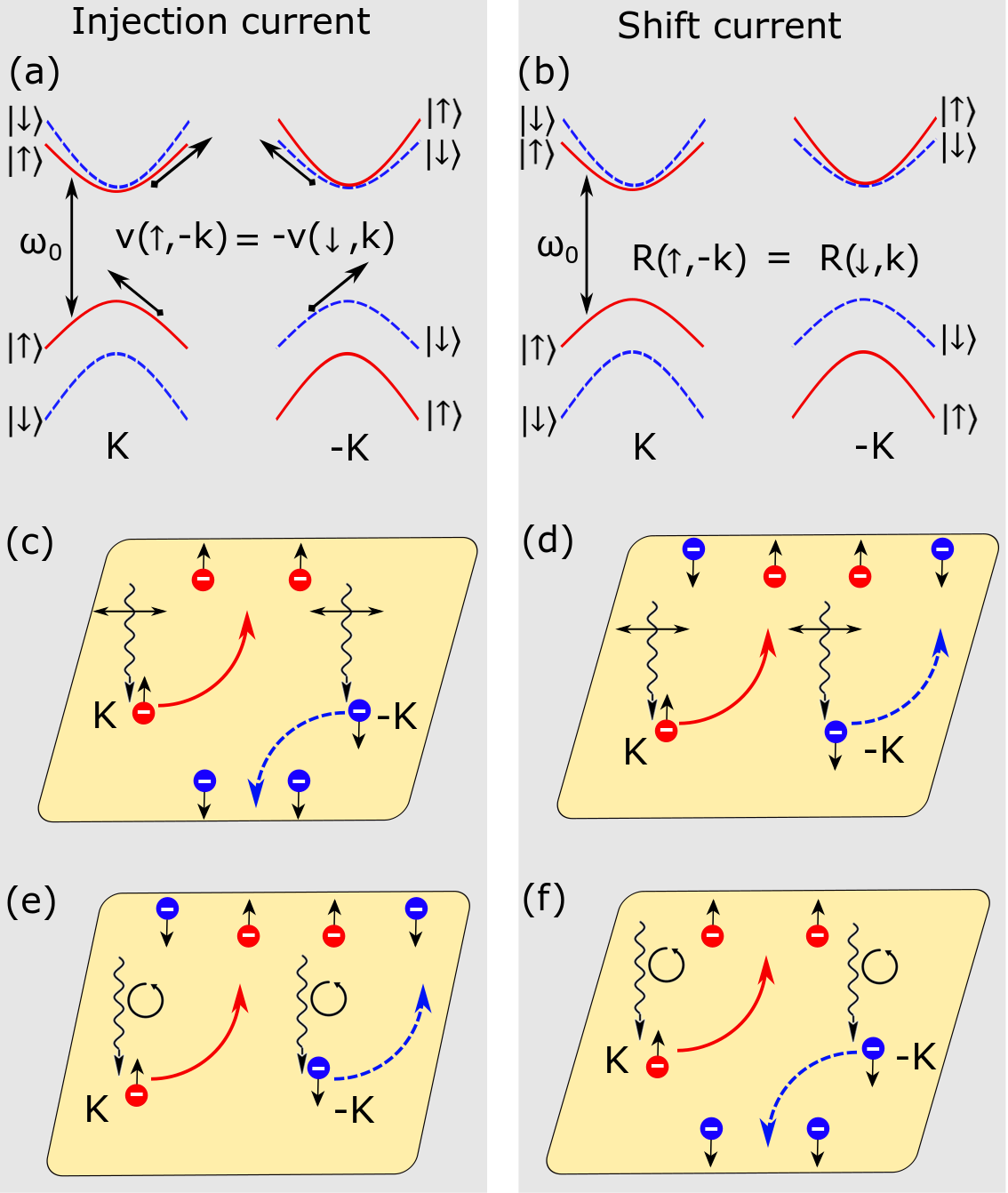}
\caption{ 
Mechanisms of the spin photogalvanic effect: 
(a) Injection-current mechanism of monolayer TMDs. 
It is decided by the group velocity, which is odd 
in reciprocal space. (b) Shift-current mechanism 
of monolayer TMDs. It is decided by the shift vector, 
which is even in reciprocal space. (c) PSC generated 
by the injection-current mechanism under linearly 
polarized light. (d) Charge current generated by 
the shift-current mechanism under linearly polarized light. 
(e) Charge current generated by the injection-current mechanism under circularly polarized light. (f) PSC generated by the shift-current mechanism under circularly polarized light.} \label{Figure1}
\end{figure}

\textit{Photocurrents under Linearly polarization:}
Using quantum perturbation theory, the DC
photoconductivity of 
injection current is (see Supplementary 
Section I \cite{Supplement} and 
refs.\cite{sipe00second,Moore_diagram} )

\begin{align}
\begin{split}
\label{eq2}
&\eta_{a b c}^{L}=\frac{-\pi e^{3}}{\hbar^{2} \omega^{2}} \sum_{m n} \int d^{3} k \alpha_{m n}^{a b}(k)\left(v_{m m}^{c}(k)-v_{n n}^{c}(k)\right) \\
&\qquad \qquad \qquad \qquad \qquad \cdot \tau \delta\left(\omega-\omega_{m n}\right)
\end{split}
\end{align}
where $\alpha_{mn}^{ab}(k)=\frac{1}{2}(v_{m n}^{a}(k) v_{n m}^{b}(k)+v_{m n}^{b}(k) v_{n m}^{a}(k)) $ 
is the optical oscillate strength.
$v_{mn}^{a}(k)$ and $v_{nm}^{b}(k)$ are the
$a$-direction and $b$-direction velocity matrices
between the conduction band $m$ and the valence
band $n$, respectively. $v_{mm}^{c}(k)$ is the
$c$-direction intraband velocity matrix. $\tau$ 
is the relaxation time of free carriers. In
time-reversal invariant materials, the optical
oscillate strength $\alpha_{mn}^{ab}(k)$ is even, 
while the term $v_{mm}^{c}(k)-v_{nn}^{c}(k)$ 
describing the group-velocity difference 
between electrons and holes is odd in reciprocal 
space, as schematically plotted in Fig. 1a. 
As a result, Eq. 2 indicates that the overall
injection-current photo-conductivity is odd.
Given the opposite orders of spin-up and spin-down 
bands between $K$ and $-K$ valleys, if pumped 
by photons ($\omega_0$), the current direction of
spin-up electrons at the K valley is opposite to 
that of spin-down electrons at the $-K$ valley 
(Fig. 1c). Namely, linearly polarized light
excites two different spins to travel in 
opposite directions, resulting in a PSC. 

Then we turn to the discussion of shift current 
in the form of (see  Supplementary section IIA
\cite{Supplement})

\begin{align}
\begin{split}
\label{eq3}
\sigma_{a b c}^{L}=\frac{\pi e^{3}}{\hbar^{2} \omega^{2}} \sum_{m n} \int d^{3} k \alpha_{m n}^{a b}(k) \mathrm{R}_{\mathrm{nm}}^{c} \delta\left(\omega-\omega_{m n}\right)
\end{split}
\end{align}

where the so-called shift vector 
$R_{nm}^{c}=\frac{\partial\phi_{nm}^{b}}{\partial k_{c}}-(A_{nn}^{c}(k)-A_{mm}^{c}(k))$ is the 
difference between Berry connections of the 
conduction band ($A_{m m}^{c}(k)$) and the valence 
band ($A_{nn}^{c}(k)$. $\phi_{nm}^{b}$ is the 
phase of the interband velocity 
matrix $v_{n m}^{b}(k)$.
Because of time-reversal symmetry, 
the shift vector and its photoconductivity 
are even in reciprocal 
space, as plotted in Fig. 1 (b). Thus,
for incident photons at the frequency $\omega_0$, 
the direction of spin-up current from the $K$ valley 
is the same as that of spin-down current from 
the $-K$ valley. As concluded in Fig. 1d, two 
different spin electrons travel in the same
direction, forming a non-spin-polarized charge 
current. This is the linearly photogalvanic effect 
or linearly bulk photovoltaic effect. 

\textit{Photocurrents under circular polarization:}
The circularly-polarized photoconductivity 
of injection current is
\begin{align}
\begin{split}
\label{eq4}
&\eta_{a b c}^{\circlearrowright}=\frac{2 \pi i e^{3}}{\hbar^{2} \omega^{2}} \sum_{m n} \int d^{3} k \operatorname{Im}(v_{m n}^{a}(k) v_{n m}^{b}(k))\\
&\qquad \qquad \qquad \qquad \cdot (v_{m m}^{c}(k)-v_{n n}^{c}(k))\tau \delta(\omega-\omega_{m n})
\end{split}
\end{align}

Recalling the Kubo formula of Berry curvature, 
$\Omega^{c}(k)=\frac{\operatorname{Im}(v_{mn}^{a}(k) v_{nm}^{b}(k))}{E_{m n}^{2}}$ \cite{xiao2010berry},
the parity of 
$Im(v_{mn}^{a}(k)v_{nm}^{b}(k))$ is the 
same as that of Berry curvature, which 
is odd in reciprocal space under time-reversal 
symmetry \cite{xiao12prl}. 
Given that the intraband transition
matrices ($v_{mm}^c(k)$ and $v_{nn}^c(k)$ in Eq. 4) 
are odd, the overall circularly-polarized
photoconductivity is even in reciprocal space.
This indicates that spin-up and spin-down 
electrons travel in the same direction, 
forming a non-spin-polarized charge current 
as shown in Fig. 1e. This so-called circularly
photogalvanic effect has been observed 
in monolayer WSe$_2$ \cite{yuan14cpge}.

Finally, the circularly-polarized photoconductivity
of shift current is
\begin{align}
\begin{split}
\label{eq5}
&\sigma_{a b c}^{\circlearrowright}=\frac{-2i\pi e^{3}}{\hbar^{2} \omega^{2}} \sum_{m n} \int d^{3} k \operatorname{Im}\left(v_{m n}^{a}(k) v_{n m}^{b}(k)\right) \\
&\qquad \qquad \qquad \qquad \qquad \cdot R_{\mathrm{nm}}^{c}(k) \delta(\omega-\omega_{mn})
\end{split}
\end{align}

Due to the odd parity of $Im(v_{mn}^a(k) v_{nm}^b(k))$
and even parity of shift vector $R_{nm}^c (k)$ 
in reciprocal space, the overall photoconductivity is odd. 
Thus, the light-excited spin-up and spin-down electrons 
transport in the opposite direction, forming a PSC as 
schematized in Fig. 1f. This unrevealed spin-polarized 
current may be one of the mechanisms responding for 
photo-induced spin current observed in quantum well
systems \cite{stevens03quantum,hubner03direct}.

\begin{figure}
\includegraphics[width=8.5cm]{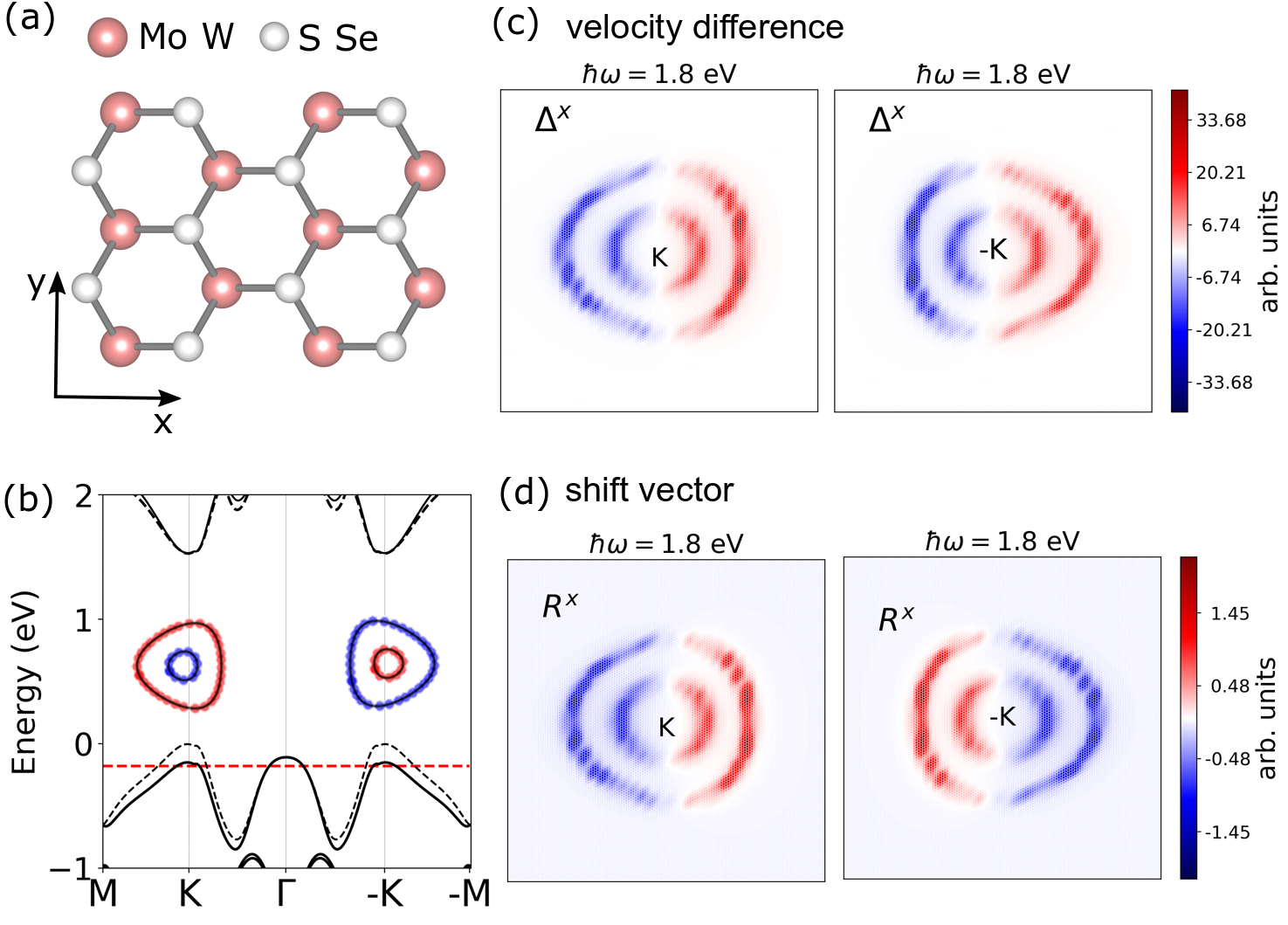}
\caption{
First-principles calculated results: (a) Top view of the
structure of monolayer group-VI dichalcogenides. 
(b) Bandstructure of monolayer MoS2. The inset is 
the calculated spin texture of electronic states
crossed by the red dashed line. Blue and red dots
represent the spin up and down along the z-direction, 
respectively.  (c) Group velocity difference 
for $\hbar\omega=1.8$ eV photon excitation. 
(d) Shift vector for $\hbar\omega=1.8$ eV photon excitation.}
\label{Fig2}
\end{figure}

\textbf{\textit{Spin and charge currents in monolayer TMDs.}}-- 
The observed spin polarization depends on the 
competition between the above-discussed currents 
in specific materials. In the following, we choose 
monolayer TMDs (Fig. 2a) as a prototypical family of 
materials for first-principles 
simulations(see details in Supplementary 
Section III \cite{Supplement,vasp1996,PAWvasp,PBE,giannozzi2009quantum}). 
The SOC and 
broken inversion symmetry enable the 
valley-spin locking in monolayer structures and 
induce the known spin and valley 
Hall effects \cite{xiao12prl} 
and circular dichroism \cite{zeng12valley,cao2012valley,mak2012control}. 
In Fig. 2b, we show the first-principles 
calculated band structure of monolayer MoS$_2$. 
The SOC splitting of the top two valence bands
is around 150 meV at $K$/$-K$ points,
which is similar to previous work \cite{liu2013three}. 
The insert plot shows the spin texture 
of electronic states around
the $K$ and $-K$ points below the top of valence bands 
(marked by a horizontal dashed line). It is 
interesting but not surprising to find that nearly 
all spins are along the z-axis, that is
$\sigma_{z}=\pm 1$,  for electrons around 
$K$ and $-K$ points \cite{saito16super}. 

As determined by Eqs. 2-5, the origins of spin and 
charge currents are mainly from parities of the 
group velocity difference $\Delta_{m n}^{c}=v_{m m}^{c}(k)-v_{n n}^{c}(k)$
and shift vector $R_{nm}^c(k)$, repectively.
Figs. 2c and 2d show these physical quantities for 
the photon energy $\hbar\omega=1.8$ eV, 
which pumps electrons
around $K$ and $-K$ points. For the sake of simplicity,
only $x$ components of these quantities are plotted. 
It is clear to see that the group velocity difference 
is even while the shift vector is odd according to 
the $\Gamma$ point of reciprocal space. This agrees
with our aforementioned discussion. 

\begin{figure}
\includegraphics[width=9cm]{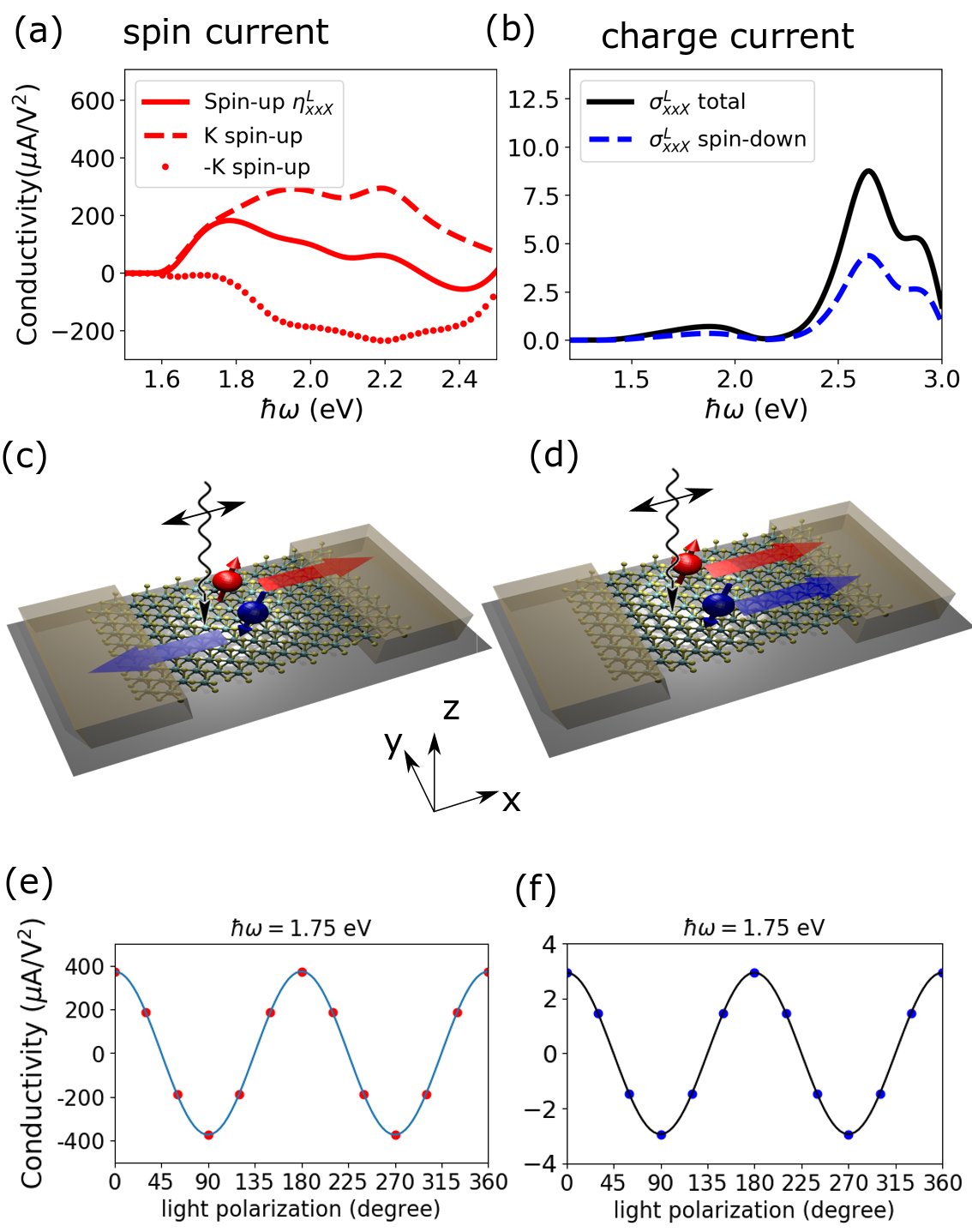}
\caption{ 
PSC under linear polarization: (a) Calculated x-direction 
spin-up current conductivity $\eta_{xxX}^{L}$ in monolayer MoS$_2$. 
(b) Charge current conductivity $\sigma_{xxX}^{L}$ 
along the $x$-direction. The $x$-direction defined in Fig. 2 (a).  
The schematic plots of light-induced the PSC (c) and charge 
current (d). (e) and (f) The light polarization direction-dependent
spin current and charge current for $\hbar\omega=1.75$ eV 
photon excitation, respectively. The zero-degree is set
to be along the x-direction.  
} 
\label{Fig3}
\end{figure}

\textit{Linear-polarization induced PSC}:
We employ the first-principles simulation \cite{Supplement}
to calculate photocurrents and photo-induced 
spin current. 
First, we focus on the injection current.
For monolayer TMDs showed in Fig. 2 (a),
only the $x$-direction inject current
is nonzero while the $y$-direction current is zero because
of the mirror symmetry with respect to the $x$-axis. 
According to Eq. 2, the carrier relaxation time $\tau$ is 
needed for obtaining quantitative results. Previous 
studies show that this relaxation time is substantially
smaller than the observed spin-relaxation time and 
electron-hole recombination
time \cite{hao16direct,wang18colloquium,wang15ultrafast}. 
Therefore, we conservatively choose $\tau=0.2$ ps 
based on results of monolayer MoS$_2$ at 
300 $K$ \cite{Kaasbjerg12first,Radisavljevic2013},
which is significantly smaller than the
estimated spin-relaxation time $\tau=40$ ps at room
temperature \cite{Yang13prl,Dankert2017}.

Fig. 3a shows the calculated spectrum of spin-up-current
conductivity excited by light polarized along 
the $x$-direction. The dashed red line is the contribution 
from the $K$ valley, and the dotted red line is that from 
the $-K$ valley. The K-valley contribution is positive 
(along the x-direction) and starts from the DFT
bandgap at 1.6 eV while the $-K$ valley contribution is 
negative and starts from $1.75$ eV because of the $150$ meV 
SOC splitting (Figure 2 (b)). Because of this competition, 
the total spin-up current (solid red line) is along the 
$x$-direction and reaches the maximum value at 1.75 eV.
Meanwhile, the light will also excite the spin-down 
current dominated by band edges at the $-K$ point 
but along the opposite ($-x$) direction, and the total
spin-down current has the same amplitude as the 
spin-up current. As a result, spin-up and spin-down 
currents are in opposite direction and result
in a zero-charge current, forming a PSC by 
linearly polarized light shown in Fig. 3c.

The linearly-polarized light will also
generate shift current, and its spectrum 
is presented in Fig. 3b. Importantly, both 
spin-up and spin-down carriers contributed 
equally to the total current along the same 
direction, as schematically shown in Fig. 3d. 
This agrees with the discussion of Eq. 3. 
The total current is a charge current but 
not spin-polarized. However, the spin current 
amplitude in Fig. 3a is about two orders of 
magnitude larger than that of the charge current 
in Fig. 3b. Therefore, the spin-polarized 
inject current will dominate observed 
photocurrent of monolayer MoS$_2$ under 
linearly polarized light, and a nearly $100\%$ PSC is expected. 

\begin{table}[!]
	\caption{\label{table_1} 
		Symmetries of interband
		Comparison of spin and charge current conductivity 
		of monolayer TMDs for linear and circular polarizations.
		We list the conductivity (in unit $\mu A/V^{2}$) at the 
		characteristic photon frequency which is marked in 
		the parentheses. This photon energy is the one starting 
		to pump both two spin electrons within a single valley, 
		e.g. 1.75 eV for monolayer MoS$_2$. }
	\begin{ruledtabular}
		\begin{tabular}{c c c c c }
			\multirow{2}{*}{Materials} & \multicolumn{2}{c}{Linear polarization}
			& \multicolumn{2}{c}{Circular polarization}  \\ \cline{2-5} 
			&\begin{tabular}[c]{@{}c@{}}Spin \\ current\\  (injection)\end{tabular} & \begin{tabular}[c]{@{}c@{}}Charge \\ current\\ (shift)\end{tabular} & \begin{tabular}[c]{@{}c@{}}Spin \\ current\\ (shift)\end{tabular} & \begin{tabular}[c]{@{}c@{}}Charge \\ current\\ (injection)\end{tabular} \\ \hline
			\begin{tabular}[c]{@{}c@{}}MoS$_2$ \\ (1.75 eV)\end{tabular} & 398 & 1   & 2   & 10 \\
			\hline
			\begin{tabular}[c]{@{}c@{}}MoSe$_2$ \\(1.50 eV)\end{tabular} & 514 & 4  & 5  & 26  \\
			\hline
			\begin{tabular}[c]{@{}c@{}}WS$_2$ \\(1.70 eV) \end{tabular}  & 556  & 2 & 3  & 8   \\
			\hline
			\begin{tabular}[c]{@{}c@{}}WSe$_2$  \\(1.65  eV) \end{tabular} & 360  & 1  & 2 & 15 \\
		\end{tabular}
	\end{ruledtabular}
\end{table}

We also calculate the spin current of the 
other important monolayer TMDs, e. g., 
MoSe$_2$, WS$_2$, and WSe$_2$. Table I summarizes
the characteristic photoconductivity caused 
by two mechanisms at photon energy that starts 
to pump the second spin within a single valley, e.g. 1.75 eV 
of monolayer MoS$_2$ in the above discussions. 
We find that monolayer WS$_2$ has the largest 
spin-polarized current, which is from its large 
SOC splitting. Although SOC splitting of monolayer 
WSe$_2$ is also large (around 0.47 eV), its group 
velocity difference $\Delta_{nm}^c$ is smaller than other 
TMDs, resulting in a smaller spin current.
Finally, many-electron effects \cite{Qiu13prl}
may quantitatively 
change these single-particle results, such as
pumping-photon energies listed in Table I.
However, without breaking any essential symmetry 
of the above discussions, many-electron effects 
will not fundamentally change our main 
conclusions of photocurrents.  

Interestingly, we find that the polarization direction
of light can switch the directions of both spin and 
charge currents. Figs. 3e and 3f show PSC and charge 
current under different polarization directions for 
photon energy at 1.75 eV, in which the $x$-direction of
MoS$_2$ (Fig. 2a) is set to be 0 degrees. The cosine 
function relations are observed for both spin and 
charge currents in the same phase. Thus, we can 
increase/decrease and even turn off the spin 
and charge currents by controlling the light 
polarization. This is from the relations between 
photoconductivities: $\eta_{xxX}^{L}=-\eta_{yyx}^{L}$
and $\sigma_{xxX}^{L}=-\sigma_{yyx}^{L}$  
enforced by the three-fold rotational symmetry of monolayer TMDs.

\textit{Spin and charge currents under circularly polarized light}:
For circularly components of photocurrents, the 
imaginary part of optical oscillation strength 
contributes to photocurrent (Eqs. 4 and 5). 
In Fig. 4a, we have plotted the calculated $Im(v_{mn}^x(k)v_{nm}^y(k))$ 
of monolayer MoS$_2$ near $K$ and $-K$ valleys excited
by photons with an energy of 1.8 eV. Our first-principles 
simulation confirms that $Im(v_{mn}^x(k)v_{nm}^y(k))$ has 
odd parity (anti-phase) in reciprocal space. As we 
discussed in Eqs. 4 and 5, 
the parity of $Im(v_{mn}^x(k)v_{nm}^y(k))$ is odd while 
the real part of optical oscillation strength is even. 
Thus, the shift-current mechanism alternatively 
generates a spin current under circularly polarized 
light as shown in Fig. 4b, in which spin-up and 
spin-down currents are in opposite direction. 
On the contrary, charge current is generated by 
the injection-current mechanism, characterized as 
the circularly photogalvanic effect. Importantly,
compare the amplitude of PSC and charge current in 
Figs. 4b and 4c, we find that the charge current 
is much stronger than spin current and the non-spin-polarized 
current is dominant in the circular polarization case. 
Table I also confirms that the charge current is about
an order of magnitude larger than the spin current 
in our calculated monolayer TMDs. Thus, we expect to
observe the circularly photogalvanic effect in monolayer 
TMDs but with a weak spin polarization [39]. 
Similarly, the charge-current direction can be 
switched by the handedness of incident light.

Finally, our predicted spin photogalvanic 
effect is universal for noncentrosymmetric 
materials, including bulk GeTe, CdS, and 2D 
group IV monochalcogenides, and it does not depend 
on the origin of SOC. We noticed that the magnitude of circularly
photogalvanic conductivity showed in Fig. 4c is 
one order smaller than that of PSC conductivity induced 
by linear polarized light. Thus, we expect that the PSC
can be measured given the fact that the 
circularly photogalvanic
effect is observed \cite{yuan14cpge}.
Our findings not only expand 
the understandings of nonlinear light-matter 
interaction physics but also substantially
broadens our capability to manipulate spintronics 
in simple nonmagnetic crystals.

\begin{figure}
\includegraphics[scale=0.23]{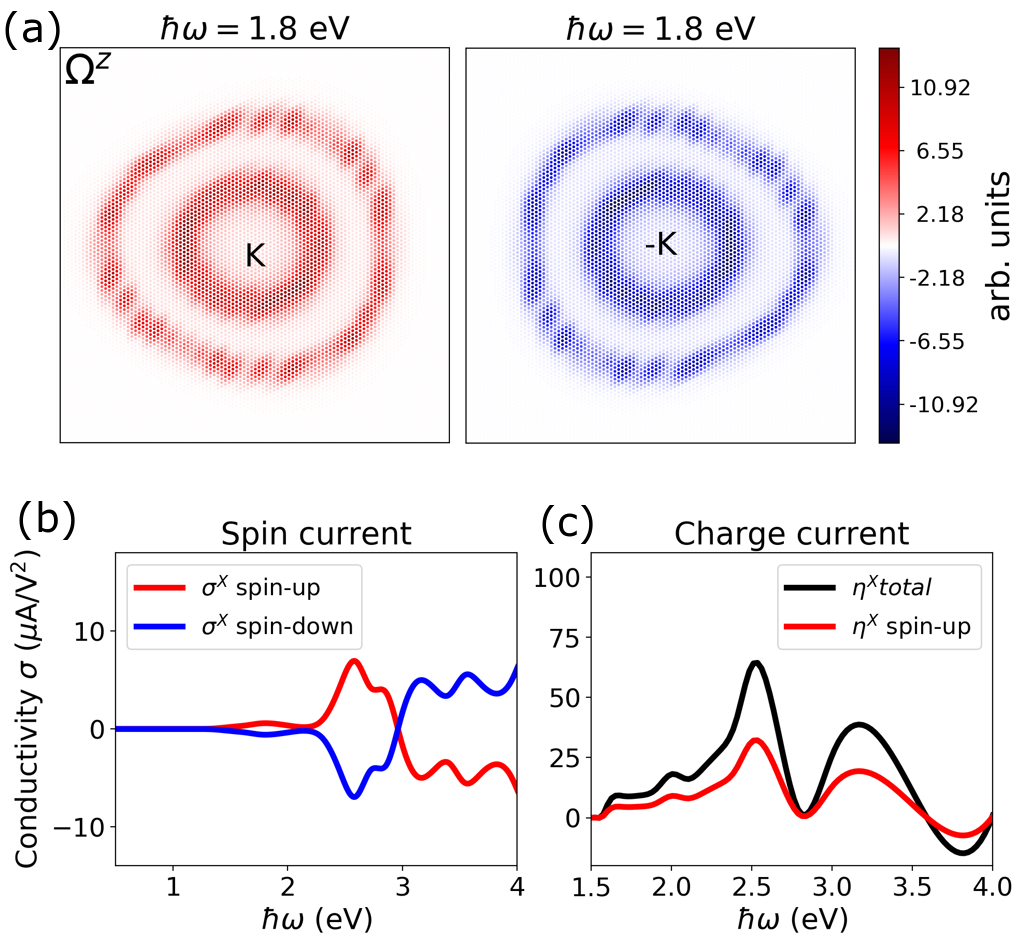}
\caption{
Spin and charge currents under circular polarization: 
(a) Imaginary part of optical transition oscillations 
(similar to the Berry curvature) for $\hbar\omega=1.8$ eV photon 
excitation. (b) PSC and (c) charge current along 
the $x$-direction for circularly polarized light.
}
\label{Fig4}
\end{figure}

\begin{acknowledgements}
This work is supported by the National 
Science Foundation (NSF) CAREER grant No. DMR-1455346, 
and the Air Force Office of Scientific Research (AFOSR) 
grant No. FA9550-17-1-0304. The computational resources
are provided by the Stampede of Teragrid at the Texas 
Advanced Computing Center (TACC) through XSEDE.
\end{acknowledgements}

\bibliography{ms}

\begin{thebibliography}{62}%
\makeatletter
\providecommand \@ifxundefined [1]{%
 \@ifx{#1\undefined}
}%
\providecommand \@ifnum [1]{%
 \ifnum #1\expandafter \@firstoftwo
 \else \expandafter \@secondoftwo
 \fi
}%
\providecommand \@ifx [1]{%
 \ifx #1\expandafter \@firstoftwo
 \else \expandafter \@secondoftwo
 \fi
}%
\providecommand \natexlab [1]{#1}%
\providecommand \enquote  [1]{``#1''}%
\providecommand \bibnamefont  [1]{#1}%
\providecommand \bibfnamefont [1]{#1}%
\providecommand \citenamefont [1]{#1}%
\providecommand \href@noop [0]{\@secondoftwo}%
\providecommand \href [0]{\begingroup \@sanitize@url \@href}%
\providecommand \@href[1]{\@@startlink{#1}\@@href}%
\providecommand \@@href[1]{\endgroup#1\@@endlink}%
\providecommand \@sanitize@url [0]{\catcode `\\12\catcode `\$12\catcode
  `\&12\catcode `\#12\catcode `\^12\catcode `\_12\catcode `\%12\relax}%
\providecommand \@@startlink[1]{}%
\providecommand \@@endlink[0]{}%
\providecommand \url  [0]{\begingroup\@sanitize@url \@url }%
\providecommand \@url [1]{\endgroup\@href {#1}{\urlprefix }}%
\providecommand \urlprefix  [0]{URL }%
\providecommand \Eprint [0]{\href }%
\providecommand \doibase [0]{http://dx.doi.org/}%
\providecommand \selectlanguage [0]{\@gobble}%
\providecommand \bibinfo  [0]{\@secondoftwo}%
\providecommand \bibfield  [0]{\@secondoftwo}%
\providecommand \translation [1]{[#1]}%
\providecommand \BibitemOpen [0]{}%
\providecommand \bibitemStop [0]{}%
\providecommand \bibitemNoStop [0]{.\EOS\space}%
\providecommand \EOS [0]{\spacefactor3000\relax}%
\providecommand \BibitemShut  [1]{\csname bibitem#1\endcsname}%
\let\auto@bib@innerbib\@empty
\bibitem [{\citenamefont {\ifmmode \check{Z}\else
  \v{Z}\fi{}uti\ifmmode~\acute{c}\else \'{c}\fi{}}\ \emph
  {et~al.}(2004)\citenamefont {\ifmmode \check{Z}\else
  \v{Z}\fi{}uti\ifmmode~\acute{c}\else \'{c}\fi{}}, \citenamefont {Fabian},\
  and\ \citenamefont {Das~Sarma}}]{sarma04RMP}%
  \BibitemOpen
  \bibfield  {author} {\bibinfo {author} {\bibfnamefont {I.}~\bibnamefont
  {\ifmmode \check{Z}\else \v{Z}\fi{}uti\ifmmode~\acute{c}\else \'{c}\fi{}}},
  \bibinfo {author} {\bibfnamefont {J.}~\bibnamefont {Fabian}}, \ and\ \bibinfo
  {author} {\bibfnamefont {S.}~\bibnamefont {Das~Sarma}},\ }\href {\doibase
  10.1103/RevModPhys.76.323} {\bibfield  {journal} {\bibinfo  {journal} {Rev.
  Mod. Phys.}\ }\textbf {\bibinfo {volume} {76}},\ \bibinfo {pages} {323}
  (\bibinfo {year} {2004})}\BibitemShut {NoStop}%
\bibitem [{\citenamefont {Bader}\ and\ \citenamefont
  {Parkin}(2010)}]{bader2010spintronics}%
  \BibitemOpen
  \bibfield  {author} {\bibinfo {author} {\bibfnamefont {S.}~\bibnamefont
  {Bader}}\ and\ \bibinfo {author} {\bibfnamefont {S.}~\bibnamefont {Parkin}},\
  }\href@noop {} {\bibfield  {journal} {\bibinfo  {journal} {Annu. Rev.
  Condens. Matter Phys.}\ }\textbf {\bibinfo {volume} {1}},\ \bibinfo {pages}
  {71} (\bibinfo {year} {2010})}\BibitemShut {NoStop}%
\bibitem [{\citenamefont {Wolf}\ \emph {et~al.}(2001)\citenamefont {Wolf},
  \citenamefont {Awschalom}, \citenamefont {Buhrman}, \citenamefont {Daughton},
  \citenamefont {von Moln{\'a}r}, \citenamefont {Roukes}, \citenamefont
  {Chtchelkanova},\ and\ \citenamefont {Treger}}]{wolf2001spintronics}%
  \BibitemOpen
  \bibfield  {author} {\bibinfo {author} {\bibfnamefont {S.}~\bibnamefont
  {Wolf}}, \bibinfo {author} {\bibfnamefont {D.}~\bibnamefont {Awschalom}},
  \bibinfo {author} {\bibfnamefont {R.}~\bibnamefont {Buhrman}}, \bibinfo
  {author} {\bibfnamefont {J.}~\bibnamefont {Daughton}}, \bibinfo {author}
  {\bibfnamefont {v.~S.}\ \bibnamefont {von Moln{\'a}r}}, \bibinfo {author}
  {\bibfnamefont {M.}~\bibnamefont {Roukes}}, \bibinfo {author} {\bibfnamefont
  {A.~Y.}\ \bibnamefont {Chtchelkanova}}, \ and\ \bibinfo {author}
  {\bibfnamefont {D.}~\bibnamefont {Treger}},\ }\href@noop {} {\bibfield
  {journal} {\bibinfo  {journal} {Science}\ }\textbf {\bibinfo {volume}
  {294}},\ \bibinfo {pages} {1488} (\bibinfo {year} {2001})}\BibitemShut
  {NoStop}%
\bibitem [{\citenamefont {Marrows}\ and\ \citenamefont
  {Dalton}(2004)}]{Marrows04prl}%
  \BibitemOpen
  \bibfield  {author} {\bibinfo {author} {\bibfnamefont {C.~H.}\ \bibnamefont
  {Marrows}}\ and\ \bibinfo {author} {\bibfnamefont {B.~C.}\ \bibnamefont
  {Dalton}},\ }\href {\doibase 10.1103/PhysRevLett.92.097206} {\bibfield
  {journal} {\bibinfo  {journal} {Physical Review Letters}\ }\textbf {\bibinfo
  {volume} {92}},\ \bibinfo {pages} {097206} (\bibinfo {year}
  {2004})}\BibitemShut {NoStop}%
\bibitem [{\citenamefont {Bhat}\ \emph {et~al.}(2005)\citenamefont {Bhat},
  \citenamefont {Nastos}, \citenamefont {Najmaie},\ and\ \citenamefont
  {Sipe}}]{bhat2005pure}%
  \BibitemOpen
  \bibfield  {author} {\bibinfo {author} {\bibfnamefont {R.~R.}\ \bibnamefont
  {Bhat}}, \bibinfo {author} {\bibfnamefont {F.}~\bibnamefont {Nastos}},
  \bibinfo {author} {\bibfnamefont {A.}~\bibnamefont {Najmaie}}, \ and\
  \bibinfo {author} {\bibfnamefont {J.}~\bibnamefont {Sipe}},\ }\href@noop {}
  {\bibfield  {journal} {\bibinfo  {journal} {Physical Review Letters}\
  }\textbf {\bibinfo {volume} {94}},\ \bibinfo {pages} {096603} (\bibinfo
  {year} {2005})}\BibitemShut {NoStop}%
\bibitem [{\citenamefont {Demidov}\ \emph {et~al.}(2012)\citenamefont
  {Demidov}, \citenamefont {Urazhdin}, \citenamefont {Ulrichs}, \citenamefont
  {Tiberkevich}, \citenamefont {Slavin}, \citenamefont {Baither}, \citenamefont
  {Schmitz},\ and\ \citenamefont {Demokritov}}]{demidov2012magnetic}%
  \BibitemOpen
  \bibfield  {author} {\bibinfo {author} {\bibfnamefont {V.~E.}\ \bibnamefont
  {Demidov}}, \bibinfo {author} {\bibfnamefont {S.}~\bibnamefont {Urazhdin}},
  \bibinfo {author} {\bibfnamefont {H.}~\bibnamefont {Ulrichs}}, \bibinfo
  {author} {\bibfnamefont {V.}~\bibnamefont {Tiberkevich}}, \bibinfo {author}
  {\bibfnamefont {A.}~\bibnamefont {Slavin}}, \bibinfo {author} {\bibfnamefont
  {D.}~\bibnamefont {Baither}}, \bibinfo {author} {\bibfnamefont
  {G.}~\bibnamefont {Schmitz}}, \ and\ \bibinfo {author} {\bibfnamefont
  {S.~O.}\ \bibnamefont {Demokritov}},\ }\href@noop {} {\bibfield  {journal}
  {\bibinfo  {journal} {Nature materials}\ }\textbf {\bibinfo {volume} {11}},\
  \bibinfo {pages} {1028} (\bibinfo {year} {2012})}\BibitemShut {NoStop}%
\bibitem [{\citenamefont {Yang}\ \emph {et~al.}(2008)\citenamefont {Yang},
  \citenamefont {Kimura},\ and\ \citenamefont {Otani}}]{yang2008giant}%
  \BibitemOpen
  \bibfield  {author} {\bibinfo {author} {\bibfnamefont {T.}~\bibnamefont
  {Yang}}, \bibinfo {author} {\bibfnamefont {T.}~\bibnamefont {Kimura}}, \ and\
  \bibinfo {author} {\bibfnamefont {Y.}~\bibnamefont {Otani}},\ }\href@noop {}
  {\bibfield  {journal} {\bibinfo  {journal} {Nature Physics}\ }\textbf
  {\bibinfo {volume} {4}},\ \bibinfo {pages} {851} (\bibinfo {year}
  {2008})}\BibitemShut {NoStop}%
\bibitem [{\citenamefont {Awschalom}\ and\ \citenamefont
  {Flatt{\'e}}(2007)}]{awschalom2007challenges}%
  \BibitemOpen
  \bibfield  {author} {\bibinfo {author} {\bibfnamefont {D.~D.}\ \bibnamefont
  {Awschalom}}\ and\ \bibinfo {author} {\bibfnamefont {M.~E.}\ \bibnamefont
  {Flatt{\'e}}},\ }\href@noop {} {\bibfield  {journal} {\bibinfo  {journal}
  {Nature Physics}\ }\textbf {\bibinfo {volume} {3}},\ \bibinfo {pages} {153}
  (\bibinfo {year} {2007})}\BibitemShut {NoStop}%
\bibitem [{\citenamefont {Urban}\ \emph {et~al.}(2001)\citenamefont {Urban},
  \citenamefont {Woltersdorf},\ and\ \citenamefont {Heinrich}}]{Gilbert01prl}%
  \BibitemOpen
  \bibfield  {author} {\bibinfo {author} {\bibfnamefont {R.}~\bibnamefont
  {Urban}}, \bibinfo {author} {\bibfnamefont {G.}~\bibnamefont {Woltersdorf}},
  \ and\ \bibinfo {author} {\bibfnamefont {B.}~\bibnamefont {Heinrich}},\
  }\href {\doibase 10.1103/PhysRevLett.87.217204} {\bibfield  {journal}
  {\bibinfo  {journal} {Physical Review Letters}\ }\textbf {\bibinfo {volume}
  {87}},\ \bibinfo {pages} {217204} (\bibinfo {year} {2001})}\BibitemShut
  {NoStop}%
\bibitem [{\citenamefont {Kajiwara}\ \emph {et~al.}(2010)\citenamefont
  {Kajiwara}, \citenamefont {Harii}, \citenamefont {Takahashi}, \citenamefont
  {Ohe}, \citenamefont {Uchida}, \citenamefont {Mizuguchi}, \citenamefont
  {Umezawa}, \citenamefont {Kawai}, \citenamefont {Ando}, \citenamefont
  {Takanashi} \emph {et~al.}}]{kajiwara10nature}%
  \BibitemOpen
  \bibfield  {author} {\bibinfo {author} {\bibfnamefont {Y.}~\bibnamefont
  {Kajiwara}}, \bibinfo {author} {\bibfnamefont {K.}~\bibnamefont {Harii}},
  \bibinfo {author} {\bibfnamefont {S.}~\bibnamefont {Takahashi}}, \bibinfo
  {author} {\bibfnamefont {J.-i.}\ \bibnamefont {Ohe}}, \bibinfo {author}
  {\bibfnamefont {K.}~\bibnamefont {Uchida}}, \bibinfo {author} {\bibfnamefont
  {M.}~\bibnamefont {Mizuguchi}}, \bibinfo {author} {\bibfnamefont
  {H.}~\bibnamefont {Umezawa}}, \bibinfo {author} {\bibfnamefont
  {H.}~\bibnamefont {Kawai}}, \bibinfo {author} {\bibfnamefont
  {K.}~\bibnamefont {Ando}}, \bibinfo {author} {\bibfnamefont {K.}~\bibnamefont
  {Takanashi}},  \emph {et~al.},\ }\href@noop {} {\bibfield  {journal}
  {\bibinfo  {journal} {Nature}\ }\textbf {\bibinfo {volume} {464}},\ \bibinfo
  {pages} {262} (\bibinfo {year} {2010})}\BibitemShut {NoStop}%
\bibitem [{\citenamefont {Sandweg}\ \emph {et~al.}(2011)\citenamefont
  {Sandweg}, \citenamefont {Kajiwara}, \citenamefont {Chumak}, \citenamefont
  {Serga}, \citenamefont {Vasyuchka}, \citenamefont {Jungfleisch},
  \citenamefont {Saitoh},\ and\ \citenamefont {Hillebrands}}]{sandweg11spin}%
  \BibitemOpen
  \bibfield  {author} {\bibinfo {author} {\bibfnamefont {C.~W.}\ \bibnamefont
  {Sandweg}}, \bibinfo {author} {\bibfnamefont {Y.}~\bibnamefont {Kajiwara}},
  \bibinfo {author} {\bibfnamefont {A.~V.}\ \bibnamefont {Chumak}}, \bibinfo
  {author} {\bibfnamefont {A.~A.}\ \bibnamefont {Serga}}, \bibinfo {author}
  {\bibfnamefont {V.~I.}\ \bibnamefont {Vasyuchka}}, \bibinfo {author}
  {\bibfnamefont {M.~B.}\ \bibnamefont {Jungfleisch}}, \bibinfo {author}
  {\bibfnamefont {E.}~\bibnamefont {Saitoh}}, \ and\ \bibinfo {author}
  {\bibfnamefont {B.}~\bibnamefont {Hillebrands}},\ }\href@noop {} {\bibfield
  {journal} {\bibinfo  {journal} {Physical Review Letters}\ }\textbf {\bibinfo
  {volume} {106}},\ \bibinfo {pages} {216601} (\bibinfo {year}
  {2011})}\BibitemShut {NoStop}%
\bibitem [{\citenamefont {Kimel}\ \emph {et~al.}(2004)\citenamefont {Kimel},
  \citenamefont {Kirilyuk}, \citenamefont {Tsvetkov}, \citenamefont {Pisarev},\
  and\ \citenamefont {Rasing}}]{kimel04laser}%
  \BibitemOpen
  \bibfield  {author} {\bibinfo {author} {\bibfnamefont {A.}~\bibnamefont
  {Kimel}}, \bibinfo {author} {\bibfnamefont {A.}~\bibnamefont {Kirilyuk}},
  \bibinfo {author} {\bibfnamefont {A.}~\bibnamefont {Tsvetkov}}, \bibinfo
  {author} {\bibfnamefont {R.}~\bibnamefont {Pisarev}}, \ and\ \bibinfo
  {author} {\bibfnamefont {T.}~\bibnamefont {Rasing}},\ }\href@noop {}
  {\bibfield  {journal} {\bibinfo  {journal} {Nature}\ }\textbf {\bibinfo
  {volume} {429}},\ \bibinfo {pages} {850} (\bibinfo {year}
  {2004})}\BibitemShut {NoStop}%
\bibitem [{\citenamefont {Uchida}\ \emph {et~al.}(2008)\citenamefont {Uchida},
  \citenamefont {Takahashi}, \citenamefont {Harii}, \citenamefont {Ieda},
  \citenamefont {Koshibae}, \citenamefont {Ando}, \citenamefont {Maekawa},\
  and\ \citenamefont {Saitoh}}]{uchida2008seebeck}%
  \BibitemOpen
  \bibfield  {author} {\bibinfo {author} {\bibfnamefont {K.}~\bibnamefont
  {Uchida}}, \bibinfo {author} {\bibfnamefont {S.}~\bibnamefont {Takahashi}},
  \bibinfo {author} {\bibfnamefont {K.}~\bibnamefont {Harii}}, \bibinfo
  {author} {\bibfnamefont {J.}~\bibnamefont {Ieda}}, \bibinfo {author}
  {\bibfnamefont {W.}~\bibnamefont {Koshibae}}, \bibinfo {author}
  {\bibfnamefont {K.}~\bibnamefont {Ando}}, \bibinfo {author} {\bibfnamefont
  {S.}~\bibnamefont {Maekawa}}, \ and\ \bibinfo {author} {\bibfnamefont
  {E.}~\bibnamefont {Saitoh}},\ }\href@noop {} {\bibfield  {journal} {\bibinfo
  {journal} {Nature}\ }\textbf {\bibinfo {volume} {455}},\ \bibinfo {pages}
  {778} (\bibinfo {year} {2008})}\BibitemShut {NoStop}%
\bibitem [{\citenamefont {Jaworski}\ \emph {et~al.}(2010)\citenamefont
  {Jaworski}, \citenamefont {Yang}, \citenamefont {Mack}, \citenamefont
  {Awschalom}, \citenamefont {Heremans},\ and\ \citenamefont
  {Myers}}]{jaworski2010seebeck}%
  \BibitemOpen
  \bibfield  {author} {\bibinfo {author} {\bibfnamefont {C.}~\bibnamefont
  {Jaworski}}, \bibinfo {author} {\bibfnamefont {J.}~\bibnamefont {Yang}},
  \bibinfo {author} {\bibfnamefont {S.}~\bibnamefont {Mack}}, \bibinfo {author}
  {\bibfnamefont {D.}~\bibnamefont {Awschalom}}, \bibinfo {author}
  {\bibfnamefont {J.}~\bibnamefont {Heremans}}, \ and\ \bibinfo {author}
  {\bibfnamefont {R.}~\bibnamefont {Myers}},\ }\href@noop {} {\bibfield
  {journal} {\bibinfo  {journal} {Nature Materials}\ }\textbf {\bibinfo
  {volume} {9}},\ \bibinfo {pages} {898} (\bibinfo {year} {2010})}\BibitemShut
  {NoStop}%
\bibitem [{\citenamefont {Qu}\ \emph {et~al.}(2013)\citenamefont {Qu},
  \citenamefont {Huang}, \citenamefont {Hu}, \citenamefont {Wu},\ and\
  \citenamefont {Chien}}]{qu13Seebeck}%
  \BibitemOpen
  \bibfield  {author} {\bibinfo {author} {\bibfnamefont {D.}~\bibnamefont
  {Qu}}, \bibinfo {author} {\bibfnamefont {S.}~\bibnamefont {Huang}}, \bibinfo
  {author} {\bibfnamefont {J.}~\bibnamefont {Hu}}, \bibinfo {author}
  {\bibfnamefont {R.}~\bibnamefont {Wu}}, \ and\ \bibinfo {author}
  {\bibfnamefont {C.}~\bibnamefont {Chien}},\ }\href@noop {} {\bibfield
  {journal} {\bibinfo  {journal} {Physical Review Letters}\ }\textbf {\bibinfo
  {volume} {110}},\ \bibinfo {pages} {067206} (\bibinfo {year}
  {2013})}\BibitemShut {NoStop}%
\bibitem [{\citenamefont {Meyer}\ \emph {et~al.}(2017)\citenamefont {Meyer},
  \citenamefont {Chen}, \citenamefont {Wimmer}, \citenamefont {Althammer},
  \citenamefont {Wimmer}, \citenamefont {Schlitz}, \citenamefont {Gepr{\"a}gs},
  \citenamefont {Huebl}, \citenamefont {K{\"o}dderitzsch}, \citenamefont
  {Ebert} \emph {et~al.}}]{meyer17nernst}%
  \BibitemOpen
  \bibfield  {author} {\bibinfo {author} {\bibfnamefont {S.}~\bibnamefont
  {Meyer}}, \bibinfo {author} {\bibfnamefont {Y.-T.}\ \bibnamefont {Chen}},
  \bibinfo {author} {\bibfnamefont {S.}~\bibnamefont {Wimmer}}, \bibinfo
  {author} {\bibfnamefont {M.}~\bibnamefont {Althammer}}, \bibinfo {author}
  {\bibfnamefont {T.}~\bibnamefont {Wimmer}}, \bibinfo {author} {\bibfnamefont
  {R.}~\bibnamefont {Schlitz}}, \bibinfo {author} {\bibfnamefont
  {S.}~\bibnamefont {Gepr{\"a}gs}}, \bibinfo {author} {\bibfnamefont
  {H.}~\bibnamefont {Huebl}}, \bibinfo {author} {\bibfnamefont
  {D.}~\bibnamefont {K{\"o}dderitzsch}}, \bibinfo {author} {\bibfnamefont
  {H.}~\bibnamefont {Ebert}},  \emph {et~al.},\ }\href@noop {} {\bibfield
  {journal} {\bibinfo  {journal} {Nature Materials}\ }\textbf {\bibinfo
  {volume} {16}},\ \bibinfo {pages} {977} (\bibinfo {year} {2017})}\BibitemShut
  {NoStop}%
\bibitem [{\citenamefont {Sheng}\ \emph {et~al.}(2017)\citenamefont {Sheng},
  \citenamefont {Sakuraba}, \citenamefont {Lau}, \citenamefont {Takahashi},
  \citenamefont {Mitani},\ and\ \citenamefont {Hayashi}}]{sheng17Nernst}%
  \BibitemOpen
  \bibfield  {author} {\bibinfo {author} {\bibfnamefont {P.}~\bibnamefont
  {Sheng}}, \bibinfo {author} {\bibfnamefont {Y.}~\bibnamefont {Sakuraba}},
  \bibinfo {author} {\bibfnamefont {Y.-C.}\ \bibnamefont {Lau}}, \bibinfo
  {author} {\bibfnamefont {S.}~\bibnamefont {Takahashi}}, \bibinfo {author}
  {\bibfnamefont {S.}~\bibnamefont {Mitani}}, \ and\ \bibinfo {author}
  {\bibfnamefont {M.}~\bibnamefont {Hayashi}},\ }\href@noop {} {\bibfield
  {journal} {\bibinfo  {journal} {Science Advances}\ }\textbf {\bibinfo
  {volume} {3}},\ \bibinfo {pages} {e1701503} (\bibinfo {year}
  {2017})}\BibitemShut {NoStop}%
\bibitem [{\citenamefont {Dyakonov}\ and\ \citenamefont
  {Perel}(1971)}]{dyakonov71current}%
  \BibitemOpen
  \bibfield  {author} {\bibinfo {author} {\bibfnamefont {M.~I.}\ \bibnamefont
  {Dyakonov}}\ and\ \bibinfo {author} {\bibfnamefont {V.}~\bibnamefont
  {Perel}},\ }\href@noop {} {\bibfield  {journal} {\bibinfo  {journal} {Physics
  Letters A}\ }\textbf {\bibinfo {volume} {35}},\ \bibinfo {pages} {459}
  (\bibinfo {year} {1971})}\BibitemShut {NoStop}%
\bibitem [{\citenamefont {Hirsch}(1999)}]{hirsch99spin}%
  \BibitemOpen
  \bibfield  {author} {\bibinfo {author} {\bibfnamefont {J.}~\bibnamefont
  {Hirsch}},\ }\href@noop {} {\bibfield  {journal} {\bibinfo  {journal}
  {Physical Review Letters}\ }\textbf {\bibinfo {volume} {83}},\ \bibinfo
  {pages} {1834} (\bibinfo {year} {1999})}\BibitemShut {NoStop}%
\bibitem [{\citenamefont {Kato}\ \emph {et~al.}(2004)\citenamefont {Kato},
  \citenamefont {Myers}, \citenamefont {Gossard},\ and\ \citenamefont
  {Awschalom}}]{kato04observation}%
  \BibitemOpen
  \bibfield  {author} {\bibinfo {author} {\bibfnamefont {Y.~K.}\ \bibnamefont
  {Kato}}, \bibinfo {author} {\bibfnamefont {R.~C.}\ \bibnamefont {Myers}},
  \bibinfo {author} {\bibfnamefont {A.~C.}\ \bibnamefont {Gossard}}, \ and\
  \bibinfo {author} {\bibfnamefont {D.~D.}\ \bibnamefont {Awschalom}},\
  }\href@noop {} {\bibfield  {journal} {\bibinfo  {journal} {Science}\ }\textbf
  {\bibinfo {volume} {306}},\ \bibinfo {pages} {1910} (\bibinfo {year}
  {2004})}\BibitemShut {NoStop}%
\bibitem [{\citenamefont {Wunderlich}\ \emph {et~al.}(2005)\citenamefont
  {Wunderlich}, \citenamefont {Kaestner}, \citenamefont {Sinova},\ and\
  \citenamefont {Jungwirth}}]{wunderlich05exper}%
  \BibitemOpen
  \bibfield  {author} {\bibinfo {author} {\bibfnamefont {J.}~\bibnamefont
  {Wunderlich}}, \bibinfo {author} {\bibfnamefont {B.}~\bibnamefont
  {Kaestner}}, \bibinfo {author} {\bibfnamefont {J.}~\bibnamefont {Sinova}}, \
  and\ \bibinfo {author} {\bibfnamefont {T.}~\bibnamefont {Jungwirth}},\
  }\href@noop {} {\bibfield  {journal} {\bibinfo  {journal} {Physical Review
  Letters}\ }\textbf {\bibinfo {volume} {94}},\ \bibinfo {pages} {047204}
  (\bibinfo {year} {2005})}\BibitemShut {NoStop}%
\bibitem [{\citenamefont {Huang}\ \emph {et~al.}(2017)\citenamefont {Huang},
  \citenamefont {Jin}, \citenamefont {Cui}, \citenamefont {Zhai}, \citenamefont
  {Mei},\ and\ \citenamefont {Liu}}]{huang17bending}%
  \BibitemOpen
  \bibfield  {author} {\bibinfo {author} {\bibfnamefont {B.}~\bibnamefont
  {Huang}}, \bibinfo {author} {\bibfnamefont {K.-H.}\ \bibnamefont {Jin}},
  \bibinfo {author} {\bibfnamefont {B.}~\bibnamefont {Cui}}, \bibinfo {author}
  {\bibfnamefont {F.}~\bibnamefont {Zhai}}, \bibinfo {author} {\bibfnamefont
  {J.}~\bibnamefont {Mei}}, \ and\ \bibinfo {author} {\bibfnamefont
  {F.}~\bibnamefont {Liu}},\ }\href@noop {} {\bibfield  {journal} {\bibinfo
  {journal} {Nature Communications}\ }\textbf {\bibinfo {volume} {8}},\
  \bibinfo {pages} {1} (\bibinfo {year} {2017})}\BibitemShut {NoStop}%
\bibitem [{\citenamefont {Murakami}\ \emph {et~al.}(2003)\citenamefont
  {Murakami}, \citenamefont {Nagaosa},\ and\ \citenamefont
  {Zhang}}]{murakami03hall}%
  \BibitemOpen
  \bibfield  {author} {\bibinfo {author} {\bibfnamefont {S.}~\bibnamefont
  {Murakami}}, \bibinfo {author} {\bibfnamefont {N.}~\bibnamefont {Nagaosa}}, \
  and\ \bibinfo {author} {\bibfnamefont {S.-C.}\ \bibnamefont {Zhang}},\
  }\href@noop {} {\bibfield  {journal} {\bibinfo  {journal} {Science}\ }\textbf
  {\bibinfo {volume} {301}},\ \bibinfo {pages} {1348} (\bibinfo {year}
  {2003})}\BibitemShut {NoStop}%
\bibitem [{\citenamefont {Sinova}\ \emph {et~al.}(2004)\citenamefont {Sinova},
  \citenamefont {Culcer}, \citenamefont {Niu}, \citenamefont {Sinitsyn},
  \citenamefont {Jungwirth},\ and\ \citenamefont {MacDonald}}]{sinova04prl}%
  \BibitemOpen
  \bibfield  {author} {\bibinfo {author} {\bibfnamefont {J.}~\bibnamefont
  {Sinova}}, \bibinfo {author} {\bibfnamefont {D.}~\bibnamefont {Culcer}},
  \bibinfo {author} {\bibfnamefont {Q.}~\bibnamefont {Niu}}, \bibinfo {author}
  {\bibfnamefont {N.}~\bibnamefont {Sinitsyn}}, \bibinfo {author}
  {\bibfnamefont {T.}~\bibnamefont {Jungwirth}}, \ and\ \bibinfo {author}
  {\bibfnamefont {A.~H.}\ \bibnamefont {MacDonald}},\ }\href@noop {} {\bibfield
   {journal} {\bibinfo  {journal} {Physical Review Letters}\ }\textbf {\bibinfo
  {volume} {92}},\ \bibinfo {pages} {126603} (\bibinfo {year}
  {2004})}\BibitemShut {NoStop}%
\bibitem [{\citenamefont {Xiao}\ \emph {et~al.}(2012)\citenamefont {Xiao},
  \citenamefont {Liu}, \citenamefont {Feng}, \citenamefont {Xu},\ and\
  \citenamefont {Yao}}]{xiao12prl}%
  \BibitemOpen
  \bibfield  {author} {\bibinfo {author} {\bibfnamefont {D.}~\bibnamefont
  {Xiao}}, \bibinfo {author} {\bibfnamefont {G.-B.}\ \bibnamefont {Liu}},
  \bibinfo {author} {\bibfnamefont {W.}~\bibnamefont {Feng}}, \bibinfo {author}
  {\bibfnamefont {X.}~\bibnamefont {Xu}}, \ and\ \bibinfo {author}
  {\bibfnamefont {W.}~\bibnamefont {Yao}},\ }\href@noop {} {\bibfield
  {journal} {\bibinfo  {journal} {Physical Review Letters}\ }\textbf {\bibinfo
  {volume} {108}},\ \bibinfo {pages} {196802} (\bibinfo {year}
  {2012})}\BibitemShut {NoStop}%
\bibitem [{\citenamefont {Ganichev}\ \emph {et~al.}(2001)\citenamefont
  {Ganichev}, \citenamefont {Ivchenko}, \citenamefont {Danilov}, \citenamefont
  {Eroms}, \citenamefont {Wegscheider}, \citenamefont {Weiss},\ and\
  \citenamefont {Prettl}}]{spin01prl}%
  \BibitemOpen
  \bibfield  {author} {\bibinfo {author} {\bibfnamefont {S.~D.}\ \bibnamefont
  {Ganichev}}, \bibinfo {author} {\bibfnamefont {E.~L.}\ \bibnamefont
  {Ivchenko}}, \bibinfo {author} {\bibfnamefont {S.~N.}\ \bibnamefont
  {Danilov}}, \bibinfo {author} {\bibfnamefont {J.}~\bibnamefont {Eroms}},
  \bibinfo {author} {\bibfnamefont {W.}~\bibnamefont {Wegscheider}}, \bibinfo
  {author} {\bibfnamefont {D.}~\bibnamefont {Weiss}}, \ and\ \bibinfo {author}
  {\bibfnamefont {W.}~\bibnamefont {Prettl}},\ }\href {\doibase
  10.1103/PhysRevLett.86.4358} {\bibfield  {journal} {\bibinfo  {journal}
  {Physical Review Letters}\ }\textbf {\bibinfo {volume} {86}},\ \bibinfo
  {pages} {4358} (\bibinfo {year} {2001})}\BibitemShut {NoStop}%
\bibitem [{\citenamefont {Ganichev}\ \emph {et~al.}(2002)\citenamefont
  {Ganichev}, \citenamefont {Ivchenko}, \citenamefont {Bel'Kov}, \citenamefont
  {Tarasenko}, \citenamefont {Sollinger}, \citenamefont {Weiss}, \citenamefont
  {Wegscheider},\ and\ \citenamefont {Prettl}}]{ganichev02spin}%
  \BibitemOpen
  \bibfield  {author} {\bibinfo {author} {\bibfnamefont {S.}~\bibnamefont
  {Ganichev}}, \bibinfo {author} {\bibfnamefont {E.}~\bibnamefont {Ivchenko}},
  \bibinfo {author} {\bibfnamefont {V.}~\bibnamefont {Bel'Kov}}, \bibinfo
  {author} {\bibfnamefont {S.}~\bibnamefont {Tarasenko}}, \bibinfo {author}
  {\bibfnamefont {M.}~\bibnamefont {Sollinger}}, \bibinfo {author}
  {\bibfnamefont {D.}~\bibnamefont {Weiss}}, \bibinfo {author} {\bibfnamefont
  {W.}~\bibnamefont {Wegscheider}}, \ and\ \bibinfo {author} {\bibfnamefont
  {W.}~\bibnamefont {Prettl}},\ }\href@noop {} {\bibfield  {journal} {\bibinfo
  {journal} {Nature}\ }\textbf {\bibinfo {volume} {417}},\ \bibinfo {pages}
  {153} (\bibinfo {year} {2002})}\BibitemShut {NoStop}%
\bibitem [{\citenamefont {Stevens}\ \emph {et~al.}(2003)\citenamefont
  {Stevens}, \citenamefont {Smirl}, \citenamefont {Bhat}, \citenamefont
  {Najmaie}, \citenamefont {Sipe},\ and\ \citenamefont
  {Van~Driel}}]{stevens03quantum}%
  \BibitemOpen
  \bibfield  {author} {\bibinfo {author} {\bibfnamefont {M.~J.}\ \bibnamefont
  {Stevens}}, \bibinfo {author} {\bibfnamefont {A.~L.}\ \bibnamefont {Smirl}},
  \bibinfo {author} {\bibfnamefont {R.}~\bibnamefont {Bhat}}, \bibinfo {author}
  {\bibfnamefont {A.}~\bibnamefont {Najmaie}}, \bibinfo {author} {\bibfnamefont
  {J.}~\bibnamefont {Sipe}}, \ and\ \bibinfo {author} {\bibfnamefont
  {H.}~\bibnamefont {Van~Driel}},\ }\href@noop {} {\bibfield  {journal}
  {\bibinfo  {journal} {Physical Review Letters}\ }\textbf {\bibinfo {volume}
  {90}},\ \bibinfo {pages} {136603} (\bibinfo {year} {2003})}\BibitemShut
  {NoStop}%
\bibitem [{\citenamefont {H{\"u}bner}\ \emph {et~al.}(2003)\citenamefont
  {H{\"u}bner}, \citenamefont {R{\"u}hle}, \citenamefont {Klude}, \citenamefont
  {Hommel}, \citenamefont {Bhat}, \citenamefont {Sipe},\ and\ \citenamefont
  {Van~Driel}}]{hubner03direct}%
  \BibitemOpen
  \bibfield  {author} {\bibinfo {author} {\bibfnamefont {J.}~\bibnamefont
  {H{\"u}bner}}, \bibinfo {author} {\bibfnamefont {W.}~\bibnamefont
  {R{\"u}hle}}, \bibinfo {author} {\bibfnamefont {M.}~\bibnamefont {Klude}},
  \bibinfo {author} {\bibfnamefont {D.}~\bibnamefont {Hommel}}, \bibinfo
  {author} {\bibfnamefont {R.}~\bibnamefont {Bhat}}, \bibinfo {author}
  {\bibfnamefont {J.}~\bibnamefont {Sipe}}, \ and\ \bibinfo {author}
  {\bibfnamefont {H.}~\bibnamefont {Van~Driel}},\ }\href@noop {} {\bibfield
  {journal} {\bibinfo  {journal} {Physical Review Letters}\ }\textbf {\bibinfo
  {volume} {90}},\ \bibinfo {pages} {216601} (\bibinfo {year}
  {2003})}\BibitemShut {NoStop}%
\bibitem [{\citenamefont {Ellsworth}\ \emph {et~al.}(2016)\citenamefont
  {Ellsworth}, \citenamefont {Lu}, \citenamefont {Lan}, \citenamefont {Chang},
  \citenamefont {Li}, \citenamefont {Wang}, \citenamefont {Hu}, \citenamefont
  {Johnson}, \citenamefont {Bian}, \citenamefont {Xiao} \emph
  {et~al.}}]{ellsworth16photo}%
  \BibitemOpen
  \bibfield  {author} {\bibinfo {author} {\bibfnamefont {D.}~\bibnamefont
  {Ellsworth}}, \bibinfo {author} {\bibfnamefont {L.}~\bibnamefont {Lu}},
  \bibinfo {author} {\bibfnamefont {J.}~\bibnamefont {Lan}}, \bibinfo {author}
  {\bibfnamefont {H.}~\bibnamefont {Chang}}, \bibinfo {author} {\bibfnamefont
  {P.}~\bibnamefont {Li}}, \bibinfo {author} {\bibfnamefont {Z.}~\bibnamefont
  {Wang}}, \bibinfo {author} {\bibfnamefont {J.}~\bibnamefont {Hu}}, \bibinfo
  {author} {\bibfnamefont {B.}~\bibnamefont {Johnson}}, \bibinfo {author}
  {\bibfnamefont {Y.}~\bibnamefont {Bian}}, \bibinfo {author} {\bibfnamefont
  {J.}~\bibnamefont {Xiao}},  \emph {et~al.},\ }\href@noop {} {\bibfield
  {journal} {\bibinfo  {journal} {Nature Physics}\ }\textbf {\bibinfo {volume}
  {12}},\ \bibinfo {pages} {861} (\bibinfo {year} {2016})}\BibitemShut
  {NoStop}%
\bibitem [{\citenamefont {von Baltz}\ and\ \citenamefont
  {Kraut}(1981)}]{von81prb}%
  \BibitemOpen
  \bibfield  {author} {\bibinfo {author} {\bibfnamefont {R.}~\bibnamefont {von
  Baltz}}\ and\ \bibinfo {author} {\bibfnamefont {W.}~\bibnamefont {Kraut}},\
  }\href@noop {} {\bibfield  {journal} {\bibinfo  {journal} {Physical Review
  B}\ }\textbf {\bibinfo {volume} {23}},\ \bibinfo {pages} {5590} (\bibinfo
  {year} {1981})}\BibitemShut {NoStop}%
\bibitem [{\citenamefont {Young}\ \emph {et~al.}(2012)\citenamefont {Young},
  \citenamefont {Zheng},\ and\ \citenamefont {Rappe}}]{young12first}%
  \BibitemOpen
  \bibfield  {author} {\bibinfo {author} {\bibfnamefont {S.~M.}\ \bibnamefont
  {Young}}, \bibinfo {author} {\bibfnamefont {F.}~\bibnamefont {Zheng}}, \ and\
  \bibinfo {author} {\bibfnamefont {A.~M.}\ \bibnamefont {Rappe}},\ }\href@noop
  {} {\bibfield  {journal} {\bibinfo  {journal} {Physical review letters}\
  }\textbf {\bibinfo {volume} {109}},\ \bibinfo {pages} {236601} (\bibinfo
  {year} {2012})}\BibitemShut {NoStop}%
\bibitem [{\citenamefont {Sipe}\ and\ \citenamefont
  {Shkrebtii}(2000)}]{sipe00second}%
  \BibitemOpen
  \bibfield  {author} {\bibinfo {author} {\bibfnamefont {J.}~\bibnamefont
  {Sipe}}\ and\ \bibinfo {author} {\bibfnamefont {A.}~\bibnamefont
  {Shkrebtii}},\ }\href@noop {} {\bibfield  {journal} {\bibinfo  {journal}
  {Physical Review B}\ }\textbf {\bibinfo {volume} {61}},\ \bibinfo {pages}
  {5337} (\bibinfo {year} {2000})}\BibitemShut {NoStop}%
\bibitem [{\citenamefont {Morimoto}\ \emph {et~al.}(2016)\citenamefont
  {Morimoto}, \citenamefont {Zhong}, \citenamefont {Orenstein},\ and\
  \citenamefont {Moore}}]{morimoto16prb}%
  \BibitemOpen
  \bibfield  {author} {\bibinfo {author} {\bibfnamefont {T.}~\bibnamefont
  {Morimoto}}, \bibinfo {author} {\bibfnamefont {S.}~\bibnamefont {Zhong}},
  \bibinfo {author} {\bibfnamefont {J.}~\bibnamefont {Orenstein}}, \ and\
  \bibinfo {author} {\bibfnamefont {J.~E.}\ \bibnamefont {Moore}},\ }\href@noop
  {} {\bibfield  {journal} {\bibinfo  {journal} {Physical Review B}\ }\textbf
  {\bibinfo {volume} {94}},\ \bibinfo {pages} {245121} (\bibinfo {year}
  {2016})}\BibitemShut {NoStop}%
\bibitem [{\citenamefont {de~Juan}\ \emph {et~al.}(2017)\citenamefont
  {de~Juan}, \citenamefont {Grushin}, \citenamefont {Morimoto},\ and\
  \citenamefont {Moore}}]{de2017quantized}%
  \BibitemOpen
  \bibfield  {author} {\bibinfo {author} {\bibfnamefont {F.}~\bibnamefont
  {de~Juan}}, \bibinfo {author} {\bibfnamefont {A.~G.}\ \bibnamefont
  {Grushin}}, \bibinfo {author} {\bibfnamefont {T.}~\bibnamefont {Morimoto}}, \
  and\ \bibinfo {author} {\bibfnamefont {J.~E.}\ \bibnamefont {Moore}},\
  }\href@noop {} {\bibfield  {journal} {\bibinfo  {journal} {Nature
  Communications}\ }\textbf {\bibinfo {volume} {8}},\ \bibinfo {pages} {1}
  (\bibinfo {year} {2017})}\BibitemShut {NoStop}%
\bibitem [{\citenamefont {Sodemann}\ and\ \citenamefont
  {Fu}(2015)}]{sodemann15quantum}%
  \BibitemOpen
  \bibfield  {author} {\bibinfo {author} {\bibfnamefont {I.}~\bibnamefont
  {Sodemann}}\ and\ \bibinfo {author} {\bibfnamefont {L.}~\bibnamefont {Fu}},\
  }\href@noop {} {\bibfield  {journal} {\bibinfo  {journal} {Physical Review
  Letters}\ }\textbf {\bibinfo {volume} {115}},\ \bibinfo {pages} {216806}
  (\bibinfo {year} {2015})}\BibitemShut {NoStop}%
\bibitem [{\citenamefont {Ma}\ \emph {et~al.}(2019)\citenamefont {Ma},
  \citenamefont {Xu}, \citenamefont {Shen}, \citenamefont {MacNeill},
  \citenamefont {Fatemi}, \citenamefont {Chang}, \citenamefont {Valdivia},
  \citenamefont {Wu}, \citenamefont {Du}, \citenamefont {Hsu} \emph
  {et~al.}}]{ma19Nhall}%
  \BibitemOpen
  \bibfield  {author} {\bibinfo {author} {\bibfnamefont {Q.}~\bibnamefont
  {Ma}}, \bibinfo {author} {\bibfnamefont {S.-Y.}\ \bibnamefont {Xu}}, \bibinfo
  {author} {\bibfnamefont {H.}~\bibnamefont {Shen}}, \bibinfo {author}
  {\bibfnamefont {D.}~\bibnamefont {MacNeill}}, \bibinfo {author}
  {\bibfnamefont {V.}~\bibnamefont {Fatemi}}, \bibinfo {author} {\bibfnamefont
  {T.-R.}\ \bibnamefont {Chang}}, \bibinfo {author} {\bibfnamefont {A.~M.~M.}\
  \bibnamefont {Valdivia}}, \bibinfo {author} {\bibfnamefont {S.}~\bibnamefont
  {Wu}}, \bibinfo {author} {\bibfnamefont {Z.}~\bibnamefont {Du}}, \bibinfo
  {author} {\bibfnamefont {C.-H.}\ \bibnamefont {Hsu}},  \emph {et~al.},\
  }\href@noop {} {\bibfield  {journal} {\bibinfo  {journal} {Nature}\ }\textbf
  {\bibinfo {volume} {565}},\ \bibinfo {pages} {337} (\bibinfo {year}
  {2019})}\BibitemShut {NoStop}%
\bibitem [{\citenamefont {Kang}\ \emph {et~al.}(2019)\citenamefont {Kang},
  \citenamefont {Li}, \citenamefont {Sohn}, \citenamefont {Shan},\ and\
  \citenamefont {Mak}}]{kang19nonlinear}%
  \BibitemOpen
  \bibfield  {author} {\bibinfo {author} {\bibfnamefont {K.}~\bibnamefont
  {Kang}}, \bibinfo {author} {\bibfnamefont {T.}~\bibnamefont {Li}}, \bibinfo
  {author} {\bibfnamefont {E.}~\bibnamefont {Sohn}}, \bibinfo {author}
  {\bibfnamefont {J.}~\bibnamefont {Shan}}, \ and\ \bibinfo {author}
  {\bibfnamefont {K.~F.}\ \bibnamefont {Mak}},\ }\href@noop {} {\bibfield
  {journal} {\bibinfo  {journal} {Nature Materials}\ }\textbf {\bibinfo
  {volume} {18}},\ \bibinfo {pages} {324} (\bibinfo {year} {2019})}\BibitemShut
  {NoStop}%
\bibitem [{\citenamefont {Shao}\ \emph {et~al.}(2020)\citenamefont {Shao},
  \citenamefont {Zhang}, \citenamefont {Gurung}, \citenamefont {Yang},\ and\
  \citenamefont {Tsymbal}}]{Shao20prl}%
  \BibitemOpen
  \bibfield  {author} {\bibinfo {author} {\bibfnamefont {D.-F.}\ \bibnamefont
  {Shao}}, \bibinfo {author} {\bibfnamefont {S.-H.}\ \bibnamefont {Zhang}},
  \bibinfo {author} {\bibfnamefont {G.}~\bibnamefont {Gurung}}, \bibinfo
  {author} {\bibfnamefont {W.}~\bibnamefont {Yang}}, \ and\ \bibinfo {author}
  {\bibfnamefont {E.~Y.}\ \bibnamefont {Tsymbal}},\ }\href {\doibase
  10.1103/PhysRevLett.124.067203} {\bibfield  {journal} {\bibinfo  {journal}
  {Physical Review Letters}\ }\textbf {\bibinfo {volume} {124}},\ \bibinfo
  {pages} {067203} (\bibinfo {year} {2020})}\BibitemShut {NoStop}%
\bibitem [{\citenamefont {Wang}\ and\ \citenamefont {Qian}(2019)}]{wang19Hall}%
  \BibitemOpen
  \bibfield  {author} {\bibinfo {author} {\bibfnamefont {H.}~\bibnamefont
  {Wang}}\ and\ \bibinfo {author} {\bibfnamefont {X.}~\bibnamefont {Qian}},\
  }\href@noop {} {\bibfield  {journal} {\bibinfo  {journal} {npj Computational
  Materials}\ }\textbf {\bibinfo {volume} {5}},\ \bibinfo {pages} {1} (\bibinfo
  {year} {2019})}\BibitemShut {NoStop}%
\bibitem [{\citenamefont {Singh}\ \emph {et~al.}(2020)\citenamefont {Singh},
  \citenamefont {Kim}, \citenamefont {Rabe},\ and\ \citenamefont
  {Vanderbilt}}]{singh20arxiv}%
  \BibitemOpen
  \bibfield  {author} {\bibinfo {author} {\bibfnamefont {S.}~\bibnamefont
  {Singh}}, \bibinfo {author} {\bibfnamefont {J.}~\bibnamefont {Kim}}, \bibinfo
  {author} {\bibfnamefont {K.~M.}\ \bibnamefont {Rabe}}, \ and\ \bibinfo
  {author} {\bibfnamefont {D.}~\bibnamefont {Vanderbilt}},\ }\href@noop {}
  {\bibfield  {journal} {\bibinfo  {journal} {arXiv preprint arXiv:2001.08283}\
  } (\bibinfo {year} {2020})}\BibitemShut {NoStop}%
\bibitem [{Sup()}]{Supplement}%
  \BibitemOpen
  \href@noop {} {\enquote {\bibinfo {title} {{S}ee {S}upplemental {M}aterial at
  xx, which includes {R}efs. [46-49]. {W}e show derivations of both injection
  and shift current under linear and circular polarization using {F}eynman
  diagrams approach, the fisrt-principles calculation details, and},}\
  }\BibitemShut {NoStop}%
\bibitem [{\citenamefont {Parker}\ \emph {et~al.}(2019)\citenamefont {Parker},
  \citenamefont {Morimoto}, \citenamefont {Orenstein},\ and\ \citenamefont
  {Moore}}]{Moore_diagram}%
  \BibitemOpen
  \bibfield  {author} {\bibinfo {author} {\bibfnamefont {D.~E.}\ \bibnamefont
  {Parker}}, \bibinfo {author} {\bibfnamefont {T.}~\bibnamefont {Morimoto}},
  \bibinfo {author} {\bibfnamefont {J.}~\bibnamefont {Orenstein}}, \ and\
  \bibinfo {author} {\bibfnamefont {J.~E.}\ \bibnamefont {Moore}},\ }\href
  {\doibase 10.1103/PhysRevB.99.045121} {\bibfield  {journal} {\bibinfo
  {journal} {Phys. Rev. B}\ }\textbf {\bibinfo {volume} {99}},\ \bibinfo
  {pages} {045121} (\bibinfo {year} {2019})}\BibitemShut {NoStop}%
\bibitem [{\citenamefont {Xiao}\ \emph {et~al.}(2010)\citenamefont {Xiao},
  \citenamefont {Chang},\ and\ \citenamefont {Niu}}]{xiao2010berry}%
  \BibitemOpen
  \bibfield  {author} {\bibinfo {author} {\bibfnamefont {D.}~\bibnamefont
  {Xiao}}, \bibinfo {author} {\bibfnamefont {M.-C.}\ \bibnamefont {Chang}}, \
  and\ \bibinfo {author} {\bibfnamefont {Q.}~\bibnamefont {Niu}},\ }\href@noop
  {} {\bibfield  {journal} {\bibinfo  {journal} {Reviews of Modern Physics}\
  }\textbf {\bibinfo {volume} {82}},\ \bibinfo {pages} {1959} (\bibinfo {year}
  {2010})}\BibitemShut {NoStop}%
\bibitem [{\citenamefont {Yuan}\ \emph {et~al.}(2014)\citenamefont {Yuan},
  \citenamefont {Wang}, \citenamefont {Lian}, \citenamefont {Zhang},
  \citenamefont {Fang}, \citenamefont {Shen}, \citenamefont {Xu}, \citenamefont
  {Xu}, \citenamefont {Zhang}, \citenamefont {Hwang} \emph
  {et~al.}}]{yuan14cpge}%
  \BibitemOpen
  \bibfield  {author} {\bibinfo {author} {\bibfnamefont {H.}~\bibnamefont
  {Yuan}}, \bibinfo {author} {\bibfnamefont {X.}~\bibnamefont {Wang}}, \bibinfo
  {author} {\bibfnamefont {B.}~\bibnamefont {Lian}}, \bibinfo {author}
  {\bibfnamefont {H.}~\bibnamefont {Zhang}}, \bibinfo {author} {\bibfnamefont
  {X.}~\bibnamefont {Fang}}, \bibinfo {author} {\bibfnamefont {B.}~\bibnamefont
  {Shen}}, \bibinfo {author} {\bibfnamefont {G.}~\bibnamefont {Xu}}, \bibinfo
  {author} {\bibfnamefont {Y.}~\bibnamefont {Xu}}, \bibinfo {author}
  {\bibfnamefont {S.-C.}\ \bibnamefont {Zhang}}, \bibinfo {author}
  {\bibfnamefont {H.~Y.}\ \bibnamefont {Hwang}},  \emph {et~al.},\ }\href@noop
  {} {\bibfield  {journal} {\bibinfo  {journal} {Nature Nanotechnology}\
  }\textbf {\bibinfo {volume} {9}},\ \bibinfo {pages} {851} (\bibinfo {year}
  {2014})}\BibitemShut {NoStop}%
\bibitem [{\citenamefont {Kresse}\ and\ \citenamefont
  {Furthm\"uller}(1996)}]{vasp1996}%
  \BibitemOpen
  \bibfield  {author} {\bibinfo {author} {\bibfnamefont {G.}~\bibnamefont
  {Kresse}}\ and\ \bibinfo {author} {\bibfnamefont {J.}~\bibnamefont
  {Furthm\"uller}},\ }\href {\doibase 10.1103/PhysRevB.54.11169} {\bibfield
  {journal} {\bibinfo  {journal} {Phys. Rev. B}\ }\textbf {\bibinfo {volume}
  {54}},\ \bibinfo {pages} {11169} (\bibinfo {year} {1996})}\BibitemShut
  {NoStop}%
\bibitem [{\citenamefont {Kresse}\ and\ \citenamefont
  {Joubert}(1999)}]{PAWvasp}%
  \BibitemOpen
  \bibfield  {author} {\bibinfo {author} {\bibfnamefont {G.}~\bibnamefont
  {Kresse}}\ and\ \bibinfo {author} {\bibfnamefont {D.}~\bibnamefont
  {Joubert}},\ }\href {\doibase 10.1103/PhysRevB.59.1758} {\bibfield  {journal}
  {\bibinfo  {journal} {Phys. Rev. B}\ }\textbf {\bibinfo {volume} {59}},\
  \bibinfo {pages} {1758} (\bibinfo {year} {1999})}\BibitemShut {NoStop}%
\bibitem [{\citenamefont {Perdew}\ \emph {et~al.}(1996)\citenamefont {Perdew},
  \citenamefont {Burke},\ and\ \citenamefont {Ernzerhof}}]{PBE}%
  \BibitemOpen
  \bibfield  {author} {\bibinfo {author} {\bibfnamefont {J.~P.}\ \bibnamefont
  {Perdew}}, \bibinfo {author} {\bibfnamefont {K.}~\bibnamefont {Burke}}, \
  and\ \bibinfo {author} {\bibfnamefont {M.}~\bibnamefont {Ernzerhof}},\ }\href
  {\doibase 10.1103/PhysRevLett.77.3865} {\bibfield  {journal} {\bibinfo
  {journal} {Phys. Rev. Lett.}\ }\textbf {\bibinfo {volume} {77}},\ \bibinfo
  {pages} {3865} (\bibinfo {year} {1996})}\BibitemShut {NoStop}%
\bibitem [{\citenamefont {Giannozzi}\ \emph {et~al.}(2009)\citenamefont
  {Giannozzi}, \citenamefont {Baroni}, \citenamefont {Bonini}, \citenamefont
  {Calandra}, \citenamefont {Car}, \citenamefont {Cavazzoni}, \citenamefont
  {Ceresoli}, \citenamefont {Chiarotti}, \citenamefont {Cococcioni},
  \citenamefont {Dabo} \emph {et~al.}}]{giannozzi2009quantum}%
  \BibitemOpen
  \bibfield  {author} {\bibinfo {author} {\bibfnamefont {P.}~\bibnamefont
  {Giannozzi}}, \bibinfo {author} {\bibfnamefont {S.}~\bibnamefont {Baroni}},
  \bibinfo {author} {\bibfnamefont {N.}~\bibnamefont {Bonini}}, \bibinfo
  {author} {\bibfnamefont {M.}~\bibnamefont {Calandra}}, \bibinfo {author}
  {\bibfnamefont {R.}~\bibnamefont {Car}}, \bibinfo {author} {\bibfnamefont
  {C.}~\bibnamefont {Cavazzoni}}, \bibinfo {author} {\bibfnamefont
  {D.}~\bibnamefont {Ceresoli}}, \bibinfo {author} {\bibfnamefont {G.~L.}\
  \bibnamefont {Chiarotti}}, \bibinfo {author} {\bibfnamefont {M.}~\bibnamefont
  {Cococcioni}}, \bibinfo {author} {\bibfnamefont {I.}~\bibnamefont {Dabo}},
  \emph {et~al.},\ }\href@noop {} {\bibfield  {journal} {\bibinfo  {journal}
  {Journal of physics: Condensed matter}\ }\textbf {\bibinfo {volume} {21}},\
  \bibinfo {pages} {395502} (\bibinfo {year} {2009})}\BibitemShut {NoStop}%
\bibitem [{\citenamefont {Zeng}\ \emph {et~al.}(2012)\citenamefont {Zeng},
  \citenamefont {Dai}, \citenamefont {Yao}, \citenamefont {Xiao},\ and\
  \citenamefont {Cui}}]{zeng12valley}%
  \BibitemOpen
  \bibfield  {author} {\bibinfo {author} {\bibfnamefont {H.}~\bibnamefont
  {Zeng}}, \bibinfo {author} {\bibfnamefont {J.}~\bibnamefont {Dai}}, \bibinfo
  {author} {\bibfnamefont {W.}~\bibnamefont {Yao}}, \bibinfo {author}
  {\bibfnamefont {D.}~\bibnamefont {Xiao}}, \ and\ \bibinfo {author}
  {\bibfnamefont {X.}~\bibnamefont {Cui}},\ }\href@noop {} {\bibfield
  {journal} {\bibinfo  {journal} {Nature Nanotechnology}\ }\textbf {\bibinfo
  {volume} {7}},\ \bibinfo {pages} {490} (\bibinfo {year} {2012})}\BibitemShut
  {NoStop}%
\bibitem [{\citenamefont {Cao}\ \emph {et~al.}(2012)\citenamefont {Cao},
  \citenamefont {Wang}, \citenamefont {Han}, \citenamefont {Ye}, \citenamefont
  {Zhu}, \citenamefont {Shi}, \citenamefont {Niu}, \citenamefont {Tan},
  \citenamefont {Wang}, \citenamefont {Liu} \emph {et~al.}}]{cao2012valley}%
  \BibitemOpen
  \bibfield  {author} {\bibinfo {author} {\bibfnamefont {T.}~\bibnamefont
  {Cao}}, \bibinfo {author} {\bibfnamefont {G.}~\bibnamefont {Wang}}, \bibinfo
  {author} {\bibfnamefont {W.}~\bibnamefont {Han}}, \bibinfo {author}
  {\bibfnamefont {H.}~\bibnamefont {Ye}}, \bibinfo {author} {\bibfnamefont
  {C.}~\bibnamefont {Zhu}}, \bibinfo {author} {\bibfnamefont {J.}~\bibnamefont
  {Shi}}, \bibinfo {author} {\bibfnamefont {Q.}~\bibnamefont {Niu}}, \bibinfo
  {author} {\bibfnamefont {P.}~\bibnamefont {Tan}}, \bibinfo {author}
  {\bibfnamefont {E.}~\bibnamefont {Wang}}, \bibinfo {author} {\bibfnamefont
  {B.}~\bibnamefont {Liu}},  \emph {et~al.},\ }\href@noop {} {\bibfield
  {journal} {\bibinfo  {journal} {Nature Communications}\ }\textbf {\bibinfo
  {volume} {3}},\ \bibinfo {pages} {1} (\bibinfo {year} {2012})}\BibitemShut
  {NoStop}%
\bibitem [{\citenamefont {Mak}\ \emph {et~al.}(2012)\citenamefont {Mak},
  \citenamefont {He}, \citenamefont {Shan},\ and\ \citenamefont
  {Heinz}}]{mak2012control}%
  \BibitemOpen
  \bibfield  {author} {\bibinfo {author} {\bibfnamefont {K.~F.}\ \bibnamefont
  {Mak}}, \bibinfo {author} {\bibfnamefont {K.}~\bibnamefont {He}}, \bibinfo
  {author} {\bibfnamefont {J.}~\bibnamefont {Shan}}, \ and\ \bibinfo {author}
  {\bibfnamefont {T.~F.}\ \bibnamefont {Heinz}},\ }\href@noop {} {\bibfield
  {journal} {\bibinfo  {journal} {Nature Nanotechnology}\ }\textbf {\bibinfo
  {volume} {7}},\ \bibinfo {pages} {494} (\bibinfo {year} {2012})}\BibitemShut
  {NoStop}%
\bibitem [{\citenamefont {Liu}\ \emph {et~al.}(2013)\citenamefont {Liu},
  \citenamefont {Shan}, \citenamefont {Yao}, \citenamefont {Yao},\ and\
  \citenamefont {Xiao}}]{liu2013three}%
  \BibitemOpen
  \bibfield  {author} {\bibinfo {author} {\bibfnamefont {G.-B.}\ \bibnamefont
  {Liu}}, \bibinfo {author} {\bibfnamefont {W.-Y.}\ \bibnamefont {Shan}},
  \bibinfo {author} {\bibfnamefont {Y.}~\bibnamefont {Yao}}, \bibinfo {author}
  {\bibfnamefont {W.}~\bibnamefont {Yao}}, \ and\ \bibinfo {author}
  {\bibfnamefont {D.}~\bibnamefont {Xiao}},\ }\href@noop {} {\bibfield
  {journal} {\bibinfo  {journal} {Physical Review B}\ }\textbf {\bibinfo
  {volume} {88}},\ \bibinfo {pages} {085433} (\bibinfo {year}
  {2013})}\BibitemShut {NoStop}%
\bibitem [{\citenamefont {Saito}\ \emph {et~al.}(2016)\citenamefont {Saito},
  \citenamefont {Nakamura}, \citenamefont {Bahramy}, \citenamefont {Kohama},
  \citenamefont {Ye}, \citenamefont {Kasahara}, \citenamefont {Nakagawa},
  \citenamefont {Onga}, \citenamefont {Tokunaga}, \citenamefont {Nojima} \emph
  {et~al.}}]{saito16super}%
  \BibitemOpen
  \bibfield  {author} {\bibinfo {author} {\bibfnamefont {Y.}~\bibnamefont
  {Saito}}, \bibinfo {author} {\bibfnamefont {Y.}~\bibnamefont {Nakamura}},
  \bibinfo {author} {\bibfnamefont {M.~S.}\ \bibnamefont {Bahramy}}, \bibinfo
  {author} {\bibfnamefont {Y.}~\bibnamefont {Kohama}}, \bibinfo {author}
  {\bibfnamefont {J.}~\bibnamefont {Ye}}, \bibinfo {author} {\bibfnamefont
  {Y.}~\bibnamefont {Kasahara}}, \bibinfo {author} {\bibfnamefont
  {Y.}~\bibnamefont {Nakagawa}}, \bibinfo {author} {\bibfnamefont
  {M.}~\bibnamefont {Onga}}, \bibinfo {author} {\bibfnamefont {M.}~\bibnamefont
  {Tokunaga}}, \bibinfo {author} {\bibfnamefont {T.}~\bibnamefont {Nojima}},
  \emph {et~al.},\ }\href@noop {} {\bibfield  {journal} {\bibinfo  {journal}
  {Nature Physics}\ }\textbf {\bibinfo {volume} {12}},\ \bibinfo {pages} {144}
  (\bibinfo {year} {2016})}\BibitemShut {NoStop}%
\bibitem [{\citenamefont {Hao}\ \emph {et~al.}(2016)\citenamefont {Hao},
  \citenamefont {Moody}, \citenamefont {Wu}, \citenamefont {Dass},
  \citenamefont {Xu}, \citenamefont {Chen}, \citenamefont {Sun}, \citenamefont
  {Li}, \citenamefont {Li}, \citenamefont {MacDonald} \emph
  {et~al.}}]{hao16direct}%
  \BibitemOpen
  \bibfield  {author} {\bibinfo {author} {\bibfnamefont {K.}~\bibnamefont
  {Hao}}, \bibinfo {author} {\bibfnamefont {G.}~\bibnamefont {Moody}}, \bibinfo
  {author} {\bibfnamefont {F.}~\bibnamefont {Wu}}, \bibinfo {author}
  {\bibfnamefont {C.~K.}\ \bibnamefont {Dass}}, \bibinfo {author}
  {\bibfnamefont {L.}~\bibnamefont {Xu}}, \bibinfo {author} {\bibfnamefont
  {C.-H.}\ \bibnamefont {Chen}}, \bibinfo {author} {\bibfnamefont
  {L.}~\bibnamefont {Sun}}, \bibinfo {author} {\bibfnamefont {M.-Y.}\
  \bibnamefont {Li}}, \bibinfo {author} {\bibfnamefont {L.-J.}\ \bibnamefont
  {Li}}, \bibinfo {author} {\bibfnamefont {A.~H.}\ \bibnamefont {MacDonald}},
  \emph {et~al.},\ }\href@noop {} {\bibfield  {journal} {\bibinfo  {journal}
  {Nature Physics}\ }\textbf {\bibinfo {volume} {12}},\ \bibinfo {pages} {677}
  (\bibinfo {year} {2016})}\BibitemShut {NoStop}%
\bibitem [{\citenamefont {Wang}\ \emph {et~al.}(2018)\citenamefont {Wang},
  \citenamefont {Chernikov}, \citenamefont {Glazov}, \citenamefont {Heinz},
  \citenamefont {Marie}, \citenamefont {Amand},\ and\ \citenamefont
  {Urbaszek}}]{wang18colloquium}%
  \BibitemOpen
  \bibfield  {author} {\bibinfo {author} {\bibfnamefont {G.}~\bibnamefont
  {Wang}}, \bibinfo {author} {\bibfnamefont {A.}~\bibnamefont {Chernikov}},
  \bibinfo {author} {\bibfnamefont {M.~M.}\ \bibnamefont {Glazov}}, \bibinfo
  {author} {\bibfnamefont {T.~F.}\ \bibnamefont {Heinz}}, \bibinfo {author}
  {\bibfnamefont {X.}~\bibnamefont {Marie}}, \bibinfo {author} {\bibfnamefont
  {T.}~\bibnamefont {Amand}}, \ and\ \bibinfo {author} {\bibfnamefont
  {B.}~\bibnamefont {Urbaszek}},\ }\href@noop {} {\bibfield  {journal}
  {\bibinfo  {journal} {Reviews of Modern Physics}\ }\textbf {\bibinfo {volume}
  {90}},\ \bibinfo {pages} {021001} (\bibinfo {year} {2018})}\BibitemShut
  {NoStop}%
\bibitem [{\citenamefont {Wang}\ \emph {et~al.}(2015)\citenamefont {Wang},
  \citenamefont {Zhang}, \citenamefont {Chan}, \citenamefont {Tiwari},\ and\
  \citenamefont {Rana}}]{wang15ultrafast}%
  \BibitemOpen
  \bibfield  {author} {\bibinfo {author} {\bibfnamefont {H.}~\bibnamefont
  {Wang}}, \bibinfo {author} {\bibfnamefont {C.}~\bibnamefont {Zhang}},
  \bibinfo {author} {\bibfnamefont {W.}~\bibnamefont {Chan}}, \bibinfo {author}
  {\bibfnamefont {S.}~\bibnamefont {Tiwari}}, \ and\ \bibinfo {author}
  {\bibfnamefont {F.}~\bibnamefont {Rana}},\ }\href@noop {} {\bibfield
  {journal} {\bibinfo  {journal} {Nature Communications}\ }\textbf {\bibinfo
  {volume} {6}},\ \bibinfo {pages} {1} (\bibinfo {year} {2015})}\BibitemShut
  {NoStop}%
\bibitem [{\citenamefont {Kaasbjerg}\ \emph {et~al.}(2012)\citenamefont
  {Kaasbjerg}, \citenamefont {Thygesen},\ and\ \citenamefont
  {Jacobsen}}]{Kaasbjerg12first}%
  \BibitemOpen
  \bibfield  {author} {\bibinfo {author} {\bibfnamefont {K.}~\bibnamefont
  {Kaasbjerg}}, \bibinfo {author} {\bibfnamefont {K.~S.}\ \bibnamefont
  {Thygesen}}, \ and\ \bibinfo {author} {\bibfnamefont {K.~W.}\ \bibnamefont
  {Jacobsen}},\ }\href {\doibase 10.1103/PhysRevB.85.115317} {\bibfield
  {journal} {\bibinfo  {journal} {Phys. Rev. B}\ }\textbf {\bibinfo {volume}
  {85}},\ \bibinfo {pages} {115317} (\bibinfo {year} {2012})}\BibitemShut
  {NoStop}%
\bibitem [{\citenamefont {Radisavljevic}\ and\ \citenamefont
  {Kis}(2013)}]{Radisavljevic2013}%
  \BibitemOpen
  \bibfield  {author} {\bibinfo {author} {\bibfnamefont {B.}~\bibnamefont
  {Radisavljevic}}\ and\ \bibinfo {author} {\bibfnamefont {A.}~\bibnamefont
  {Kis}},\ }\href {\doibase 10.1038/nmat3687} {\bibfield  {journal} {\bibinfo
  {journal} {Nature Materials}\ }\textbf {\bibinfo {volume} {12}},\ \bibinfo
  {pages} {815} (\bibinfo {year} {2013})}\BibitemShut {NoStop}%
\bibitem [{\citenamefont {Song}\ and\ \citenamefont {Dery}(2013)}]{Yang13prl}%
  \BibitemOpen
  \bibfield  {author} {\bibinfo {author} {\bibfnamefont {Y.}~\bibnamefont
  {Song}}\ and\ \bibinfo {author} {\bibfnamefont {H.}~\bibnamefont {Dery}},\
  }\href {\doibase 10.1103/PhysRevLett.111.026601} {\bibfield  {journal}
  {\bibinfo  {journal} {Phys. Rev. Lett.}\ }\textbf {\bibinfo {volume} {111}},\
  \bibinfo {pages} {026601} (\bibinfo {year} {2013})}\BibitemShut {NoStop}%
\bibitem [{\citenamefont {Dankert}\ and\ \citenamefont
  {Dash}(2017)}]{Dankert2017}%
  \BibitemOpen
  \bibfield  {author} {\bibinfo {author} {\bibfnamefont {A.}~\bibnamefont
  {Dankert}}\ and\ \bibinfo {author} {\bibfnamefont {S.~P.}\ \bibnamefont
  {Dash}},\ }\href {\doibase 10.1038/ncomms16093} {\bibfield  {journal}
  {\bibinfo  {journal} {Nature Communications}\ }\textbf {\bibinfo {volume}
  {8}},\ \bibinfo {pages} {16093} (\bibinfo {year} {2017})}\BibitemShut
  {NoStop}%
\bibitem [{\citenamefont {Qiu}\ \emph {et~al.}(2013)\citenamefont {Qiu},
  \citenamefont {da~Jornada},\ and\ \citenamefont {Louie}}]{Qiu13prl}%
  \BibitemOpen
  \bibfield  {author} {\bibinfo {author} {\bibfnamefont {D.~Y.}\ \bibnamefont
  {Qiu}}, \bibinfo {author} {\bibfnamefont {F.~H.}\ \bibnamefont {da~Jornada}},
  \ and\ \bibinfo {author} {\bibfnamefont {S.~G.}\ \bibnamefont {Louie}},\
  }\href {\doibase 10.1103/PhysRevLett.111.216805} {\bibfield  {journal}
  {\bibinfo  {journal} {Phys. Rev. Lett.}\ }\textbf {\bibinfo {volume} {111}},\
  \bibinfo {pages} {216805} (\bibinfo {year} {2013})}\BibitemShut {NoStop}%
\end{thebibliography}%


\begin{thebibliography}{7}%
\makeatletter
\providecommand \@ifxundefined [1]{%
 \@ifx{#1\undefined}
}%
\providecommand \@ifnum [1]{%
 \ifnum #1\expandafter \@firstoftwo
 \else \expandafter \@secondoftwo
 \fi
}%
\providecommand \@ifx [1]{%
 \ifx #1\expandafter \@firstoftwo
 \else \expandafter \@secondoftwo
 \fi
}%
\providecommand \natexlab [1]{#1}%
\providecommand \enquote  [1]{``#1''}%
\providecommand \bibnamefont  [1]{#1}%
\providecommand \bibfnamefont [1]{#1}%
\providecommand \citenamefont [1]{#1}%
\providecommand \href@noop [0]{\@secondoftwo}%
\providecommand \href [0]{\begingroup \@sanitize@url \@href}%
\providecommand \@href[1]{\@@startlink{#1}\@@href}%
\providecommand \@@href[1]{\endgroup#1\@@endlink}%
\providecommand \@sanitize@url [0]{\catcode `\\12\catcode `\$12\catcode
  `\&12\catcode `\#12\catcode `\^12\catcode `\_12\catcode `\%12\relax}%
\providecommand \@@startlink[1]{}%
\providecommand \@@endlink[0]{}%
\providecommand \url  [0]{\begingroup\@sanitize@url \@url }%
\providecommand \@url [1]{\endgroup\@href {#1}{\urlprefix }}%
\providecommand \urlprefix  [0]{URL }%
\providecommand \Eprint [0]{\href }%
\providecommand \doibase [0]{http://dx.doi.org/}%
\providecommand \selectlanguage [0]{\@gobble}%
\providecommand \bibinfo  [0]{\@secondoftwo}%
\providecommand \bibfield  [0]{\@secondoftwo}%
\providecommand \translation [1]{[#1]}%
\providecommand \BibitemOpen [0]{}%
\providecommand \bibitemStop [0]{}%
\providecommand \bibitemNoStop [0]{.\EOS\space}%
\providecommand \EOS [0]{\spacefactor3000\relax}%
\providecommand \BibitemShut  [1]{\csname bibitem#1\endcsname}%
\let\auto@bib@innerbib\@empty
\bibitem [{\citenamefont {Parker}\ \emph {et~al.}(2019)\citenamefont {Parker},
  \citenamefont {Morimoto}, \citenamefont {Orenstein},\ and\ \citenamefont
  {Moore}}]{Moore_diagram}%
  \BibitemOpen
  \bibfield  {author} {\bibinfo {author} {\bibfnamefont {D.~E.}\ \bibnamefont
  {Parker}}, \bibinfo {author} {\bibfnamefont {T.}~\bibnamefont {Morimoto}},
  \bibinfo {author} {\bibfnamefont {J.}~\bibnamefont {Orenstein}}, \ and\
  \bibinfo {author} {\bibfnamefont {J.~E.}\ \bibnamefont {Moore}},\ }\href
  {\doibase 10.1103/PhysRevB.99.045121} {\bibfield  {journal} {\bibinfo
  {journal} {Phys. Rev. B}\ }\textbf {\bibinfo {volume} {99}},\ \bibinfo
  {pages} {045121} (\bibinfo {year} {2019})}\BibitemShut {NoStop}%
\bibitem [{\citenamefont {Sipe}\ and\ \citenamefont
  {Shkrebtii}(2000)}]{sipe2000}%
  \BibitemOpen
  \bibfield  {author} {\bibinfo {author} {\bibfnamefont {J.~E.}\ \bibnamefont
  {Sipe}}\ and\ \bibinfo {author} {\bibfnamefont {A.~I.}\ \bibnamefont
  {Shkrebtii}},\ }\href {\doibase 10.1103/PhysRevB.61.5337} {\bibfield
  {journal} {\bibinfo  {journal} {Phys. Rev. B}\ }\textbf {\bibinfo {volume}
  {61}},\ \bibinfo {pages} {5337} (\bibinfo {year} {2000})}\BibitemShut
  {NoStop}%
\bibitem [{\citenamefont {Mahan}(2013)}]{mahan2013many}%
  \BibitemOpen
  \bibfield  {author} {\bibinfo {author} {\bibfnamefont {G.~D.}\ \bibnamefont
  {Mahan}},\ }\href@noop {} {\emph {\bibinfo {title} {Many-particle physics}}}\
  (\bibinfo  {publisher} {Springer, Berlin},\ \bibinfo {year}
  {2013})\BibitemShut {NoStop}%
\bibitem [{\citenamefont {Kresse}\ and\ \citenamefont
  {Furthm\"uller}(1996)}]{vasp1996}%
  \BibitemOpen
  \bibfield  {author} {\bibinfo {author} {\bibfnamefont {G.}~\bibnamefont
  {Kresse}}\ and\ \bibinfo {author} {\bibfnamefont {J.}~\bibnamefont
  {Furthm\"uller}},\ }\href {\doibase 10.1103/PhysRevB.54.11169} {\bibfield
  {journal} {\bibinfo  {journal} {Phys. Rev. B}\ }\textbf {\bibinfo {volume}
  {54}},\ \bibinfo {pages} {11169} (\bibinfo {year} {1996})}\BibitemShut
  {NoStop}%
\bibitem [{\citenamefont {Kresse}\ and\ \citenamefont
  {Joubert}(1999)}]{PAWvasp}%
  \BibitemOpen
  \bibfield  {author} {\bibinfo {author} {\bibfnamefont {G.}~\bibnamefont
  {Kresse}}\ and\ \bibinfo {author} {\bibfnamefont {D.}~\bibnamefont
  {Joubert}},\ }\href {\doibase 10.1103/PhysRevB.59.1758} {\bibfield  {journal}
  {\bibinfo  {journal} {Phys. Rev. B}\ }\textbf {\bibinfo {volume} {59}},\
  \bibinfo {pages} {1758} (\bibinfo {year} {1999})}\BibitemShut {NoStop}%
\bibitem [{\citenamefont {Perdew}\ \emph {et~al.}(1996)\citenamefont {Perdew},
  \citenamefont {Burke},\ and\ \citenamefont {Ernzerhof}}]{PBE}%
  \BibitemOpen
  \bibfield  {author} {\bibinfo {author} {\bibfnamefont {J.~P.}\ \bibnamefont
  {Perdew}}, \bibinfo {author} {\bibfnamefont {K.}~\bibnamefont {Burke}}, \
  and\ \bibinfo {author} {\bibfnamefont {M.}~\bibnamefont {Ernzerhof}},\ }\href
  {\doibase 10.1103/PhysRevLett.77.3865} {\bibfield  {journal} {\bibinfo
  {journal} {Phys. Rev. Lett.}\ }\textbf {\bibinfo {volume} {77}},\ \bibinfo
  {pages} {3865} (\bibinfo {year} {1996})}\BibitemShut {NoStop}%
\bibitem [{\citenamefont {Giannozzi}\ \emph {et~al.}(2009)\citenamefont
  {Giannozzi}, \citenamefont {Baroni}, \citenamefont {Bonini}, \citenamefont
  {Calandra}, \citenamefont {Car}, \citenamefont {Cavazzoni}, \citenamefont
  {Ceresoli}, \citenamefont {Chiarotti}, \citenamefont {Cococcioni},
  \citenamefont {Dabo} \emph {et~al.}}]{giannozzi2009quantum}%
  \BibitemOpen
  \bibfield  {author} {\bibinfo {author} {\bibfnamefont {P.}~\bibnamefont
  {Giannozzi}}, \bibinfo {author} {\bibfnamefont {S.}~\bibnamefont {Baroni}},
  \bibinfo {author} {\bibfnamefont {N.}~\bibnamefont {Bonini}}, \bibinfo
  {author} {\bibfnamefont {M.}~\bibnamefont {Calandra}}, \bibinfo {author}
  {\bibfnamefont {R.}~\bibnamefont {Car}}, \bibinfo {author} {\bibfnamefont
  {C.}~\bibnamefont {Cavazzoni}}, \bibinfo {author} {\bibfnamefont
  {D.}~\bibnamefont {Ceresoli}}, \bibinfo {author} {\bibfnamefont {G.~L.}\
  \bibnamefont {Chiarotti}}, \bibinfo {author} {\bibfnamefont {M.}~\bibnamefont
  {Cococcioni}}, \bibinfo {author} {\bibfnamefont {I.}~\bibnamefont {Dabo}},
  \emph {et~al.},\ }\href@noop {} {\bibfield  {journal} {\bibinfo  {journal}
  {Journal of physics: Condensed matter}\ }\textbf {\bibinfo {volume} {21}},\
  \bibinfo {pages} {395502} (\bibinfo {year} {2009})}\BibitemShut {NoStop}%
\end{thebibliography}%
\end{document}


\title{Supplementary Information:\\Intrinsic Spin 
Photogalvanic Effect in Nonmagnetic Insulator}
\maketitle

\section{Injection current mechanism}
\begin{figure}[!htbp]
\centering
\includegraphics[scale=0.08]{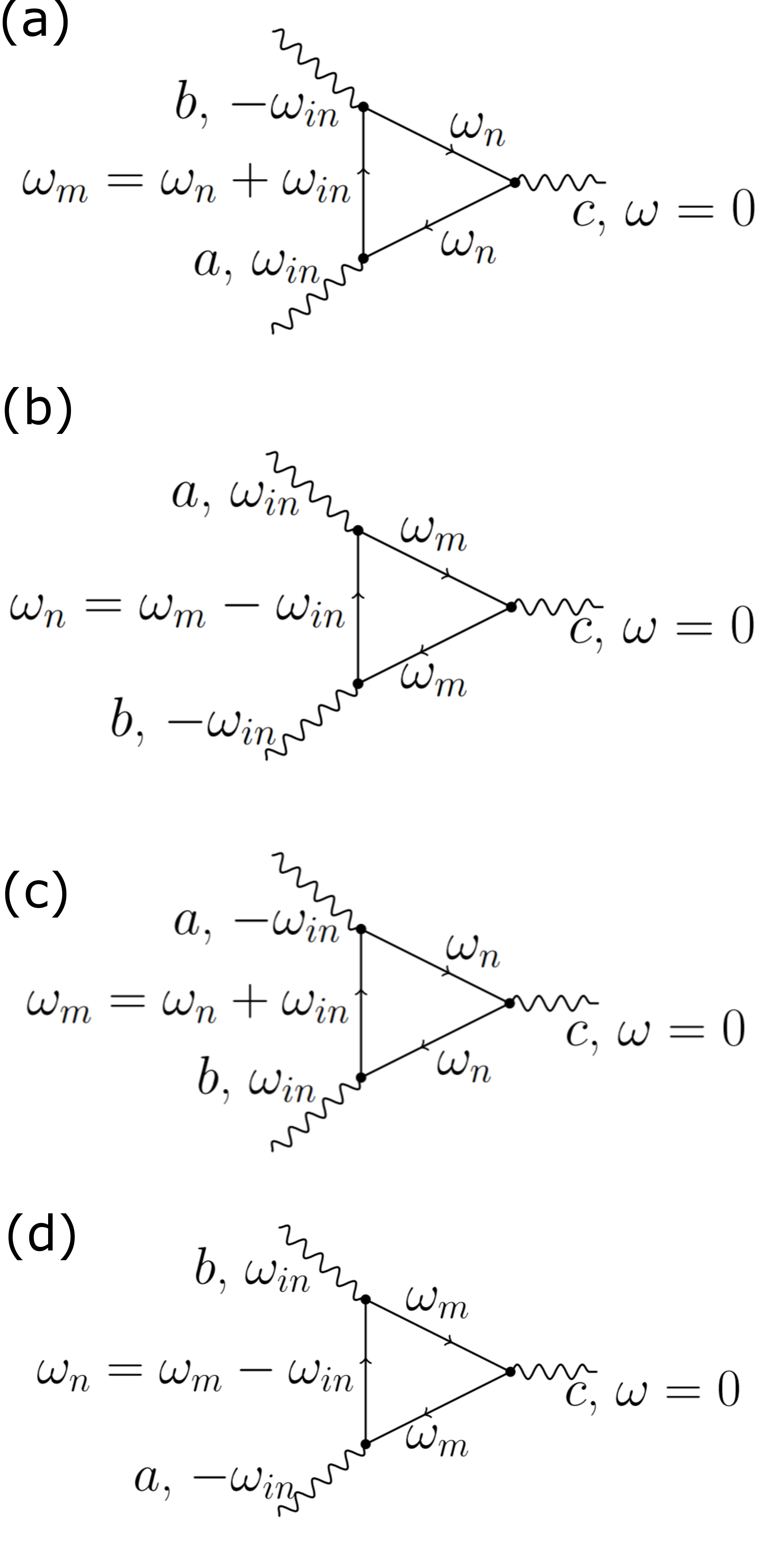}
\caption{ 
The Feynman diagram for the injection current. The difference between 
the four diagrams are the different orders of polarized
direction of light and the valence and conduction bands.
}
\label{Fig_S1} 
\end{figure}

We adapt the Feynman diagram approach to calculate 
the injection current. Using the Feynman rules derived
by D. E. Parker et.al.\cite{Moore_diagram}, 
we can easily calculate the corresponding 
injection-current tensors. Fig. \ref{Fig_S1} shows 
the Feynman diagrams of injection-current. Here, 
we use four diagrams to distinguish the order 
of polarization direction, and the conduction 
band $m$ and valence band $n$. 

For the one-photon pole  under the velocity gauge, 
it contributevelocity matrix 
element $\mel{\alpha}{\hat{v}_{a}}{\beta}$ 
for band $\alpha$ translated to any band $\beta$. Here, in the above-discussed 
diagram, we have inter-band velocity matrix element $\mel{m}{\hat{v}_{a}}{n}$ and intra-band velocity matrix element $\mel{n}{\hat{v}_{a}}{n}$ and
$\mel{m}{\hat{v}_{a}}{m}$.The occupation number of 
conduction and valence band is $f_m=0$ and $f_n=1$, respectively.

The first diagram shown by Fig. \ref{Fig_S1} (a) contribute to
\begin{align}
\begin{split}
\eta_1(0;\omega_{in}, -\omega_{in})&\\
=\frac{e^3}{\hbar^2\omega_{in}^2}\sum_{mn}&\int\frac{d^3k} {(2\pi)^3}\int{d\omega'}G_n(\omega')G_m(\omega'+\omega_{in})G_n(\omega')\\
&\cdot\mel{n}{\hat{v}_{a}}{m}\mel{m}{\hat{v}_{b}}{n} \mel{n}{\hat{v}_{c}}{n}\\
=\frac{e^3}{\hbar^2\omega_{in}^2}\sum_{mn}&\int\frac{d^3k}
{(2\pi)^3}\int{d\omega'}\frac{1}{\omega'-\omega_n+i\xi_n}\frac{1}{\omega'+\omega_{in}-\omega_m+i\xi_m}\\
&\cdot\frac{1}{\omega'-\omega_n+i\xi_n}    \mel{n}{\hat{v}_{a}}{m}\mel{m}{\hat{v}_{b}}{n} \mel{n}{\hat{v}_{c}}{n}
\label{eq1}
\end{split}
\end{align}
where the $G(\omega)$ is the free quai-particle
propagator with a dressed version. $\xi$ is 
the imaginary part of the self-energy for 
the quasi-particle.

To evaluate the integrals such as 
\begin{align}
\begin{split}
\label{eq2}
&I(\omega_{in},-\omega_{in} )\\
&=\int{d\omega'}G_n(\omega')G_m(\omega'+\omega_{in})G_n(\omega')\\
&=\int{d\omega'}\frac{1}{\omega'-\omega_n}\frac{1}{\omega'+\omega_{in}-\omega_m}\frac{1}{\omega'-\omega_n} 
\end{split}
\end{align}

We directly use the formalism deviated from the contour integral techniques with Matsubara frequencies \cite{Moore_diagram}. 
\begin{align}
\begin{split}
\label{eq3}
&I_3(\omega_{in},-\omega_{in})\\
=&\frac{(-\omega_{in}-\omega_{nm})f_{nm}+(\omega_{in}-\omega_{mn})f_{nm}}{(\omega_{in}-\omega_{mn})(-\omega_{in}-\omega_{nm})(\omega-\omega_{nn}+i\xi)}\\
=&(\frac{f_{nm}}{\omega_{in}-\omega_{mn}}+\frac{f_{nm}}{-\omega_{in}-\omega_{nm}})\frac{1}{\omega-\omega_{nn}+i\xi} \\
=&f_{nm}(\wp(\frac{1}{\omega_{in}-\omega_{mn}})+\wp(\frac{1}{-\omega_{in}-\omega_{nm}})\\
&\qquad +2i\pi\delta(\omega_{in}-\omega_{mn}))\cdot\frac{1}{\omega-\omega_{nn}+i\xi}
\end{split}
\end{align}
where the $\wp(\frac{1}{\omega_{in}-\omega_{mn}})$ 
is the principle part. $f_{nm}=f_n-f_m$ where $f_n$ 
and $f_m$ are the occupation number. Because the principle part is odd while the $\delta$ function is even, only imaginary part is non-vanishing. Thus, the first diagram contribution is 

\begin{align}
\begin{split}
\label{eq4}
&\eta_1(0;\omega_{in}, -\omega_{in})\\
&=\frac{2i\pi e^3}{\hbar^2\omega_{in}^2}\sum_{mn}\int\frac{d^3k}{(2\pi)^3} \mel{n}{\hat{v}_{a}}{m}\mel{m}{\hat{v}_{b}}{n}\\ 
&\qquad \qquad  \qquad \cdot \frac{\mel{n}{\hat{v}_{c}}{n}}{\omega-\omega_{nn}+i\xi}\delta(\omega_{in}-\omega_{mn})\\
\end{split}
\end{align}

Simarlarly, the contribution of the other three diagrams are 
\begin{align}
\begin{split}
\label{eq5}
&\eta_2(0;\omega_{in}, -\omega_{in})\\
&=\frac{2i\pi e^3}{\hbar^2\omega_{in}^2}\sum_{mn}\int\frac{d^3k}{(2\pi)^3} f_{mn}\mel{n}{\hat{v}_{a}}{m}\mel{m}{\hat{v}_{b}}{n}\\ 
&\qquad \qquad  \qquad \cdot\frac{\mel{m}{\hat{v}_{c}}{m}}{\omega-\omega_{mm}+i\xi}\delta(\omega_{in}-\omega_{mn})\\
\end{split}
\end{align}

\begin{align}
\begin{split}
\label{eq6}
&\eta_3(0;\omega_{in}, -\omega_{in})\\
&=\frac{2i\pi e^3}{\hbar^2\omega_{in}^2}\sum_{mn}\int\frac{d^3k}{(2\pi)^3} f_{nm}\mel{n}{\hat{v}_{b}}{m}\mel{m}{\hat{v}_{a}}{n}\\ 
&\qquad \qquad  \qquad \cdot\frac{\mel{n}{\hat{v}_{c}}{n}}{\omega-\omega_{nn}+i\xi}\delta(\omega_{in}-\omega_{mn})\\
\end{split}
\end{align}

\begin{align}
\begin{split}
\label{eq7}
&\eta_4(0;\omega_{in}, -\omega_{in})\\
&=\frac{2i\pi e^3}{\hbar^2\omega_{in}^2}\sum_{mn}\int\frac{d^3k}{(2\pi)^3} f_{mn}\mel{n}{\hat{v}_{b}}{m}\mel{m}{\hat{v}_{a}}{n}\\ 
&\qquad \qquad  \qquad \cdot \frac{\mel{m}{\hat{v}_{c}}{m}}{\omega-\omega_{mm}+i\xi}\delta(\omega_{in}-\omega_{mn})\\
\end{split}
\end{align}

\subsection{Linearly polarized light}
For the linearly polarized light, the current density is $j^{L}_c=\eta^{L}_{c}(E_{a}\cos(\theta)+E_{b}\sin(\theta))(E_{a}\cos(\theta)+E_{b}\sin(\theta))=\eta^{L}_{aac}E^2_{a}\cos^2(\theta)+\eta^{L}_{bbc}E^2_{bb}\sin^2(\theta)+(\eta^{L}_{abc}+\eta^{L}_{bac})E_{a}E_{b}\cos(\theta)\sin(\theta)$.

Here, we use the occupation number relation
$f_{vc}=-f_{cv}=f_v-f_c=1$. If we use the electron-hole
symmetry within the relaxation time approximation. i.e. $\tau_m=\tau_n=1/\xi\equiv \tau$. 
Then, the corresponding injection current 
conductivity tensors under linearly polarized light is 
\begin{align}
\begin{split}
\label{eq12}
&\eta^{L}_{abc}=\eta^{L}_{bac} \equiv \frac{1}{4}(\eta_1+\eta_2+\eta_3+\eta_4)=\\
&\frac{\pi e^3}{2\hbar^2\omega_{in}^2}\sum_{nm}\int\frac{d^3k}{(2\pi)^3} (\mel{n}{\hat{v}_{a}}{m}\mel{m}{\hat{v}_{b}}{n}+\mel{n}{\hat{v}_{b}}{m}\mel{m}{\hat{v}_{a}}{n})\\ 
&\qquad \qquad \qquad  \cdot(\mel{n}{\hat{v}_{c}}{n}-\mel{m}{\hat{v}_{c}}{m})\tau\delta(\omega_{in}-\omega_{mn})\\
\end{split}
\end{align}

Specially, the diagonal component tensor of the injection
current conductivity (e.g. $a=b\equiv a$) is 
\begin{multline}
\begin{split}
\label{eq13}
&\eta^{L}_{aac}\\
&=\frac{\pi e^3}{\hbar^2\omega_{in}^2}\sum_{mn}\int\frac{d^3k}{(2\pi)^3} |\mel{n}{\hat{v}_{a}}{m}|^2
(\mel{n}{\hat{v}_{c}}{n}-\mel{m}{\hat{v}_{c}}{m})\\
&\qquad \qquad \qquad  \cdot\tau\delta(\omega_{in}-\omega_{mn})\\
\end{split}
\end{multline}

\subsection{\label{sec:level2} Circularly polarized light}
Next, we derive the circular component tensor 
of the injection current conductivity.
For the time-reversal invariant non-centrosymmetric
semiconductors, the injection current is non-vanishing
under circular polarization. For right-hand circularly 
polarized light, the total current density
$j^{\circlearrowright}_{c}=\eta_{c}(E_{a}-iE_{b})(E_{a}+iE_{b})=
\eta^L_{aac}E^2_{\alpha}+\eta^L_{bbc}E^2_{\beta}+i\eta^{\circlearrowright}_{abc}E_{a}E_{b}$.
While for the left-hand circular polarized light, the current density is $j^{\circlearrowleft}_{c}=\eta_{c}(E_{a}+iE_{b})(E_{a}-iE_{b})=
\eta^L_{aac}E^2_{\alpha}+\eta^L_{bbc}E^2_{\beta}+i\eta^{\circlearrowleft}_{abc}E_{a}E_{b}$. 
For the circular component of current conductivity tensor, 
we can get the relation $\eta^{\circlearrowleft}_{abc}$=-$\eta^{\circlearrowright}_{abc}$

we also use the occupation number relation $f_{vc}=-f_{cv}=f_v-f_c=1$, and the
electron-hole symmetry within the relaxation time approximation. i.e. $\tau_c=\tau_v=1/\xi$. 
The the circular component tensor 
of the injection current conductivity is 
\begin{align}
\begin{split}
\label{eq11}
&\eta^{\circlearrowright}_{abc}=-\eta^{\circlearrowleft}_{abc}=\frac{1}{2}(\eta_1+\eta_2-\eta_3-\eta_4)\\
&=\frac{-\pi e^3}{\hbar^2\omega_{in}^2}\sum_{mn}\int\frac{d^3k}{(2\pi)^3} (\mel{n}{\hat{v}_{a}}{m}\mel{m}{\hat{v}_{b}}{n}-\mel{n}{\hat{v}_{b}}{m}\mel{m}{\hat{v}_{a}}{n})\\ 
&\qquad \qquad \qquad  \cdot(\mel{n}{\hat{v}_{c}}{n}-\mel{m}{\hat{v}_{c}}{m})\tau\delta(\omega_{in}-\omega_{mn})\\
&=\frac{-2i\pi e^3}{\hbar^2\omega_{in}^2}\sum_{mn}\int\frac{d^3k}{(2\pi)^3} Im(\mel{n}{\hat{v}_{a}}{m}\mel{m}{\hat{v}_{b}}{n})\\ 
&\qquad \qquad \qquad  \cdot(\mel{n}{\hat{v}_{c}}{n}-\mel{m}{\hat{v}_{c}}{m})\tau\delta(\omega_{in}-\omega_{mn})\\
\end{split}
\end{align}

The Eq. \ref{eq11} is same as the usually used formalism 
deviated by Sipe et. al. \cite{sipe2000} based on the polarization
operator method. We note that in Ref. \cite{sipe2000}, the 
injection current tensor $2\eta^{abc}_{2}$ is equivalent to 
the $\eta^{\circlearrowright}_{abc}$ in our case.

\section{Shift current mechanism}
In the ab-initio framework, the precision achievable for the
computation of nonlinear optical response is in general still 
poor when compared with the quality of calculated first-order
optical properties. Here, we aim to construct the
methodology of shift current nonlinear optical response
tensor. 

\begin{figure}
\centering
\includegraphics[scale=0.2]{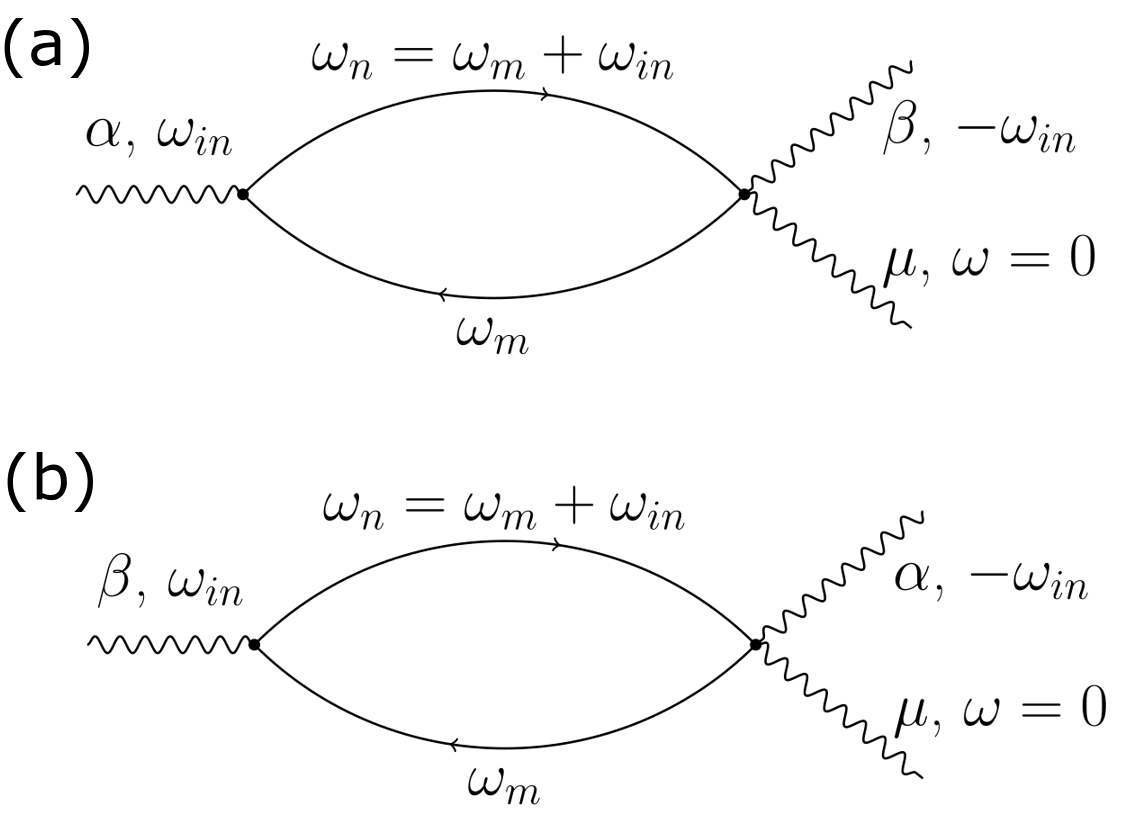}
\caption{ 
The Feynman diagram for the shift current. 
The difference between 
the four diagrams are the different orders 
of polarized direction of light.
}
\label{FigS2} 
\end{figure}

\subsection{Linearly polarized light}
The shift current conductivity tensor
at the single-particle
approximation level (shown in Fig.\ref{FigS2}) \cite{Moore_diagram} is
\begin{align}
\begin{split}
&\sigma^L_{\alpha\beta\gamma}(0;\omega_{in}, -\omega_{in})\\
&=\frac{e^3}{\hbar^2\omega_{in}^2}\sum_{mn}\int\frac{d^3k} {(2\pi)^3}\int{d\omega'}G_m(\omega')G_n(\omega'+\omega_{in})\\ 
&\qquad \cdot (\mel{m}{\hat{v}_{\alpha}}{n}\mel{n}{\hat{v}_{\beta}}{m}_{;\mu}+ \mel{m}{\hat{v}_{\beta}}{n}\mel{n}{\hat{v}_{\alpha}}{m}_{;\mu})\\
\label{eq1}
\end{split}
\end{align}

where $\mel{m}{\hat{v}_{\alpha}}{n}$ is contributed by the
one-photon vertex, and $\mel{n}{\hat{v}_{\alpha}}{m}_{;\mu}$ is
contributed by two-photon vertex using the velocity gauge. 
Here the $\mel{n}{\hat{v}_{\alpha}}{m}_{;\mu}$ is defined as
$\frac{\partial \mel{n}{\hat{v}_{\alpha}}{m}}{\partial k_{\mu}}
-i(A_{nn}^{\mu}-A_{mm}^{\mu})\mel{n}{\hat{v}_{\alpha}}{m}$,
where $A_n^{\mu}$ is the Berry connection. The $\alpha$
and $\beta$ are the polarized direction of light, $\mu$ is the current
direction.
the $G(\omega)$ is the free quai-particle propagator with a dressed
version, i.e. 
$G_m(\omega')=\frac{1}{\omega'-\omega_m-i\xi}$, 
where $\omega_m$ is electron energy at mean-field theory, $\xi$ is the imaginary part 
of quasi-particle energy. 

To evaluate the integrals such as 
\begin{align}
\begin{split}
\label{eq2}
&I(\omega_{in},-\omega_{in} )=\int{d\omega'}G_m(\omega')G_n(\omega'+\omega_{in})\\
&=\int{d\omega'}\frac{1}{\omega'-\omega_m}\cdot\frac{1}{\omega'+\omega_{in}-\omega_n}
\end{split}
\end{align}

We use the formalism deviated from the contour integral techniques with Matsubara frequencies \cite{mahan2013many}

\begin{align}
\label{eq2}
&I(\omega_{in},-\omega_{in} )=\frac{f_{mn}}{\omega_{in}-\omega_{mn}+i\xi}
\end{align}
where $f_{mn}=f_m-f_n$ is the occupation number
difference of quasi-particle states $\ket{m}$ 
and $\ket{n}$. In general, $I(\omega_{in},-\omega_{in})$
have real and imaginary part, similar to the
dielectric susceptibility.

However, the current susceptibility $\sigma(0;\omega_{in}, -\omega_{in})$ 
is real and the imaginary part is zero.
To see this, we replace the quasi-particle states 
$\ket{m}$ and $\ket{n}$ to the conduction particle 
$\ket{c}$ and valence particle $\ket{v}$.

1). If $\ket{m}=\ket{c}$ and $\ket{n}=\ket{v}$,
we have $\omega_{in}=\omega>0$; with Eq. S13, the
Eq. \ref{eq1} goes to

\begin{align}
\begin{split}
&\sigma^L_1(0; \omega, -\omega)\\
&=\frac{e^3}{\hbar^2\omega^2}\sum_{vc}\int\frac{d^3k}{(2\pi)^3}\frac{f_{cv}}{\omega-\omega_{cv}+i\xi}\\ 
&\qquad \qquad \qquad \left(\mel{c}{\hat{v}_{\alpha}}{v}\mel{v}{\hat{v}_{\beta}}{c}_{;\mu}+\mel{c}{\hat{v}_{\beta}}{v}\mel{v}{\hat{v}_{\alpha}}{c}_{;\mu}\right)\\
&=\frac{e^3}{\hbar^2\omega^2}\sum_{vc}\int\frac{d^3k} {(2\pi)^3}\frac{1}{\omega-\omega_{cv}+i\xi}\\
&\qquad \qquad \qquad \left(\hat{v}^{\alpha}_{cv}(k)\hat{v}^{\beta}_{vc;\mu}(k)+\hat{v}^{\beta}_{cv}(k)\hat{v}^{\alpha}_{vc;\mu}(k)\right)\\
\end{split}
\label{current1}
\end{align}

2). If $\ket{m}=\ket{v}$ and $\ket{n}=\ket{c}$,
we have $\omega_{in}=-\omega<0$. Then, the
Eq.\ref{eq1} goes to

\begin{align}
\begin{split}
&\sigma^L_2(0; -\omega, \omega)\\
&=\frac{e^3}{\hbar^2\omega^2}\sum_{vc}\int\frac{d^3k} {(2\pi)^3}\frac{f_{vc}}{-\omega-\omega_{vc}+i\xi}\\
&\qquad \qquad \qquad \left(\mel{v}{\hat{v}_{\alpha}}{c}\mel{c}{\hat{v}_{\beta}}{v}_{;\mu}+\mel{v}{\hat{v}_{\beta}}{c}\mel{c}{\hat{v}_{\alpha}}{v}_{;\mu}\right)\\
&=\frac{e^3}{\hbar^2\omega^2}\sum_{vc}\int\frac{d^3k} {(2\pi)^3}\frac{1}{\omega_{in}+\omega_{vc}-i\xi}\\
&\qquad \qquad \qquad \left(\hat{v}^{\alpha}_{vc}(k)\hat{v}^{\beta}_{cv;\mu}(k)+\hat{v}^{\beta}_{vc}(k)\hat{v}^{\alpha}_{cv;\mu}(k)\right)\\
\end{split}
\label{current2}
\end{align}

When the system have the time-reversal symmetry, 
we have the following relations:

\begin{align}
\begin{split}
\hat{v}^{\alpha}_{vc}(k)\frac{\partial\hat{v}^{\beta}_{cv}(k)}{\partial k_{\mu}}&=-\hat{v}^{\alpha}_{cv}(-k)\frac{\partial\hat{v}^{\beta}_{vc}(-k)}{\partial(-k_{\mu})}
\\
\hat{v}^{\alpha}_{vc}(k)\hat{v}^{\beta}_{cv}(k)&=\hat{v}^{\alpha}_{cv}(-k)\hat{v}^{\beta}_{vc}(-k)\\
i\hat{\xi}^{\mu}_{vv}(k)&=-i\hat{\xi}^{\mu}_{vv}(-k)\\
\end{split}
\label{relation}
\end{align}

Using Eq. \ref{relation} and defination of $\hat{v}^{\beta}_{cv;\mu}(k)$, 
the Eq. \ref{current2} can change to
\begin{align}
\begin{split}
&\sigma^L_2(0; -\omega, \omega)\\
&=\frac{e^3}{\hbar^2\omega^2}\sum_{vc}\int\frac{d^3k} {(2\pi)^3}\frac{1}{\omega+\omega_{vc}-i\xi}\\
&\qquad \qquad \qquad \left(-\hat{v}^{\alpha}_{cv}(-k)\hat{v}^{\beta}_{vc;\mu}(-k)-\hat{v}^{\beta}_{cv}(-k)\hat{v}^{\alpha}_{vc;\mu}(-k)\right)\\
&=\frac{e^3}{\hbar^2\omega^2}\sum_{vc}\int\frac{d^3k} {(2\pi)^3}\frac{-1}{\omega-\omega_{cv}-i\xi}\\
&\qquad \qquad \qquad \left(\hat{v}^{\alpha}_{cv}(k)\hat{v}^{\beta}_{vc;\mu}(k)+\hat{v}^{\beta}_{cv}(k)\hat{v}^{\alpha}_{vc;\mu}(k)\right)\\
\end{split}
\label{current2_mod}
\end{align}

We sum the Eq. \ref{current1} and Eq. \ref{current2_mod}, then
the shift current susceptibility or 
conductivity is 
\begin{align}
\begin{split}
&\sigma^L_{\alpha\beta\gamma}(0; \omega, -\omega)
=\frac{1}{2}\left(\sigma^L_1(0;\omega,
-\omega)+\sigma^L_2(0; -\omega, \omega)\right)\\
&=\frac{i\pi e^3}{\hbar^2\omega^2}\sum_{vc}\int\frac{d^3k} {(2\pi)^3}
\left(\hat{v}^{\alpha}_{cv}(k)\hat{v}^{\beta}_{vc;\mu}(k)+
\hat{v}^{\beta}_{cv}(k)\hat{v}^{\alpha}_{vc;\mu}(k)\right)\\
&\qquad \qquad \qquad \qquad \cdot \delta(\omega-\omega_{cv})\\
\end{split}
\label{current_total}
\end{align}

We know that the $\hat{v}^{\alpha}_{vc;\mu}=\frac{\partial \mel{v}{\hat{v}_{\alpha}}{c}}{\partial k_{\mu}}
-i(A_{vv}^{\mu}-A_{cc}^{\mu})\mel{v}{\hat{v}_{\alpha}}{c}$. Here we let $ \mel{v}{\hat{v}_{\alpha}}{c}=|\mel{v}{\hat{v}_{\alpha}}{c}|e^{i\phi^{\alpha}_{vc}}$, then $\hat{v}^{\alpha}_{vc;\mu}=\frac{\partial |\mel{v}{\hat{v}_{\alpha}}{c}|}{\partial k_{\mu}}e^{i\phi^{\alpha}_{vc}}+(i\frac{\partial \phi^{\alpha}_{vc}}{\partial k_{\mu}}
-i(A_{vv}^{\mu}-A_{cc}^{\mu}))\mel{v}{\hat{v}_{\alpha}}{c}$, where the $\frac{\partial |\mel{v}{\hat{v}_{\alpha}}{c}|}{\partial k_{\mu}}e^{i\phi^{\alpha}_{vc}}$ is odd 
in the reciprocal space. Thus, the 
the shift current conductivity is 

\begin{align}
	\begin{split}
		&\sigma^L_{\alpha\beta\gamma}(0; \omega, -\omega)=\frac{-\pi e^3}{\hbar^2\omega^2}\sum_{vc}\int\frac{d^3k} {(2\pi)^3}
		(\hat{v}^{\alpha}_{cv}(k)\hat{v}^{\beta}_{vc}(k)R^{\alpha,\mu}_{vc}(k)+\\
		&\qquad \qquad \qquad \qquad \qquad \qquad
		\hat{v}^{\beta}_{cv}(k)\hat{v}^{\alpha}_{vc}(k)R^{\beta,\mu}_{vc}(k))
		\cdot \delta(\omega-\omega_{cv})\\
	\end{split}
	\label{current_total}
\end{align}

where the shift vector $R^{\alpha,\mu}_{vc}(k)=\frac{\partial \phi^{\alpha}_{vc}}{\partial k_{\mu}}
-(A_{vv}^{\mu}-A_{cc}^{\mu})$. 
In the isotropic system, e.g. MoS$_2$, the 
the phase of the interband velocity matrix
is not direction dependent, i. e. $\phi^{\alpha}_{vc}\simeq \phi^{\beta}_{vc}$.
So Eq. S19 can be simplified as

\begin{align}
\begin{split}
&\sigma^L(0; \omega, -\omega)
=\frac{1}{2}\left(\sigma^L_1(0;\omega,
-\omega)+\sigma^L_2(0; -\omega, \omega)\right)\\
&=\frac{-\pi e^3}{\hbar^2\omega^2}\sum_{vc}\int\frac{d^3k} {(2\pi)^3}
\left(\hat{v}^{\alpha}_{cv}(k)\hat{v}^{\beta}_{vc}(k)+
\hat{v}^{\beta}_{cv}(k)\hat{v}^{\alpha}_{vc}(k)\right)\\
&\qquad \qquad \qquad \qquad \qquad R^{\mu}_{vc}(k)\cdot \delta(\omega-\omega_{cv})\\
\end{split}
\label{current_total}
\end{align}

\subsection{Circularly polarized light}
For the circular component of the shift
current conductivity tensor, the formula
is replace Eq. S11 by
\begin{align}
\begin{split}
&\sigma^{\circlearrowright}_{\alpha\beta\gamma}(0;\omega_{in}, -\omega_{in})=-\sigma^{\circlearrowleft}_{\alpha\beta\gamma}(0;\omega_{in}, -\omega_{in})\\
&=\frac{e^3}{\hbar^2\omega_{in}^2}\sum_{mn}\int\frac{d^3k} {(2\pi)^3}\int{d\omega'}G_m(\omega')G_n(\omega'+\omega_{in})\\ 
&\qquad \cdot (\mel{m}{\hat{v}_{\alpha}}{n}\mel{n}{\hat{v}_{\beta}}{m}_{;\mu}- \mel{m}{\hat{v}_{\beta}}{n}\mel{n}{\hat{v}_{\alpha}}{m}_{;\mu})\\
\label{eq1}
\end{split}
\end{align}

Proceeding the derivation similar to the 
Eq. S12-S19, the circular component of 
the shift current conductivity is 
\begin{align}
\begin{split}
&\sigma^{\circlearrowright}_{\alpha\beta\gamma}(0; \omega, -\omega)
= -\sigma^{\circlearrowleft}_{\alpha\beta\gamma}(0;\omega_{in}, -\omega_{in})\\
&=\frac{-\pi e^3}{\hbar^2\omega^2}\sum_{vc}\int\frac{d^3k} {(2\pi)^3}
\left(\hat{v}^{\alpha}_{cv}(k)\hat{v}^{\beta}_{vc}(k)-
\hat{v}^{\beta}_{cv}(k)\hat{v}^{\alpha}_{vc}(k)\right)\\
&\qquad \qquad \qquad \qquad \qquad R^{\mu}_{vc}(k)\cdot \delta(\omega-\omega_{cv})\\
\end{split}
\label{current_total}
\end{align}
 
 To consistent with the indices of injection current in our
 manuscript,  we replace the band index c and v by m and n
 respectively, the directional index $\alpha$, $\beta$, $\gamma$
 by index $a$, $b$, $c$ respectively.
 
\section{First-principles Computational details}
\begin{figure}[!htbp]
\includegraphics[scale=0.14]{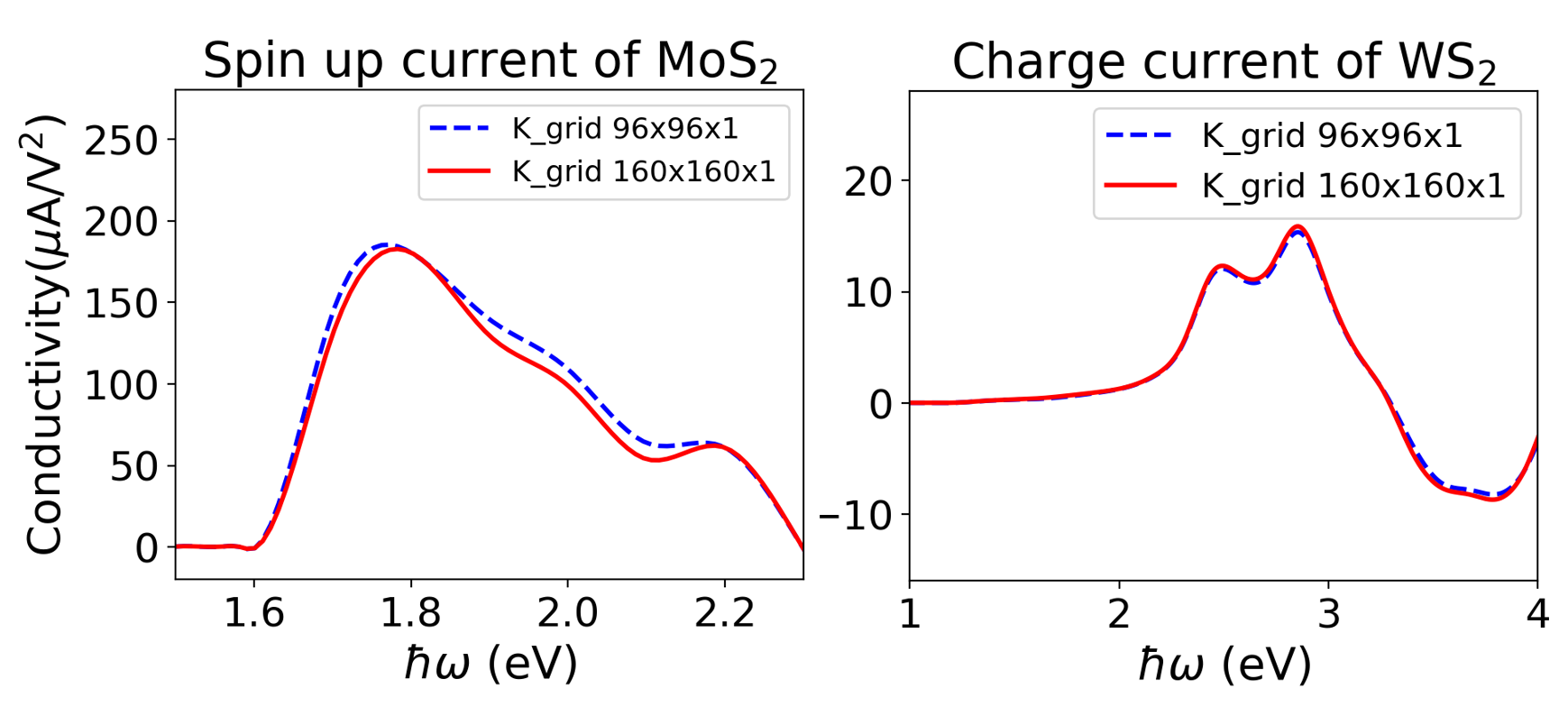}
\caption{(left)The spin up current conductivity $\eta^L_{xxX}$
of monolayer MoS$_2$ using $96\times96 \times1$ 
and $160\times160 \times1$ k-grid samples. 
(Right)The charge current conductivity $\sigma^L_{xxX}$
of monolayer MoS$_2$ using $96\times96 \times1$ 
and $160\times160 \times1$ k-grid samples.
}
\label{FigS3} 
\end{figure}

The First principles calculations such as velocity
matrix, optical oscillate strength, Berry curvature
are mainly performed 
with the Vienna Ab-initio Simulation Package
(VASP) \cite{vasp1996} using the projector augmented
wave method \cite{PAWvasp} and the plane-wave basis 
with an energy cutoff of 500 eV. The Perdew-Burke-Ernzerhof (PBE)
exchange-correlation functional \cite{PBE} was used 
with spin-orbital coupling (SOC).  The shift vector,  and
shift current conductivity are also checked using the 
Quantum ESPRESSO package\cite{giannozzi2009quantum} with 
PBE exchange-correlation functional.

The Structure optimizations were performed with a 
force criterion of $0.01$ $eV$/$\AA$. The Monkhorst-Pack 
k-point meshes of $10\times10\times1$ were adopted for the calculations of monolayer TMDs structures. 

For nonlinear photocurrent calculation, we use the 
Monkhorst-Pack k-point meshes of $160\times160 \times1$ to get the
converged spin and charge current. 
Fig. \ref{FigS3} presents the spin current conductivity
element $\eta_{xxX}^L$ of MoS$_2$ (left) and
the charge current conductivity element $\sigma_{xxX}^L$ 
of WS$_2$ (right) for different k-grid samples. 
The results using the k-point meshes of $160\times160 \times1$ 
are converged.

\section{Injection current contributions}
\subsection{Spin current under linear polarization}

\begin{figure}[!htbp]
\includegraphics[scale=0.22]{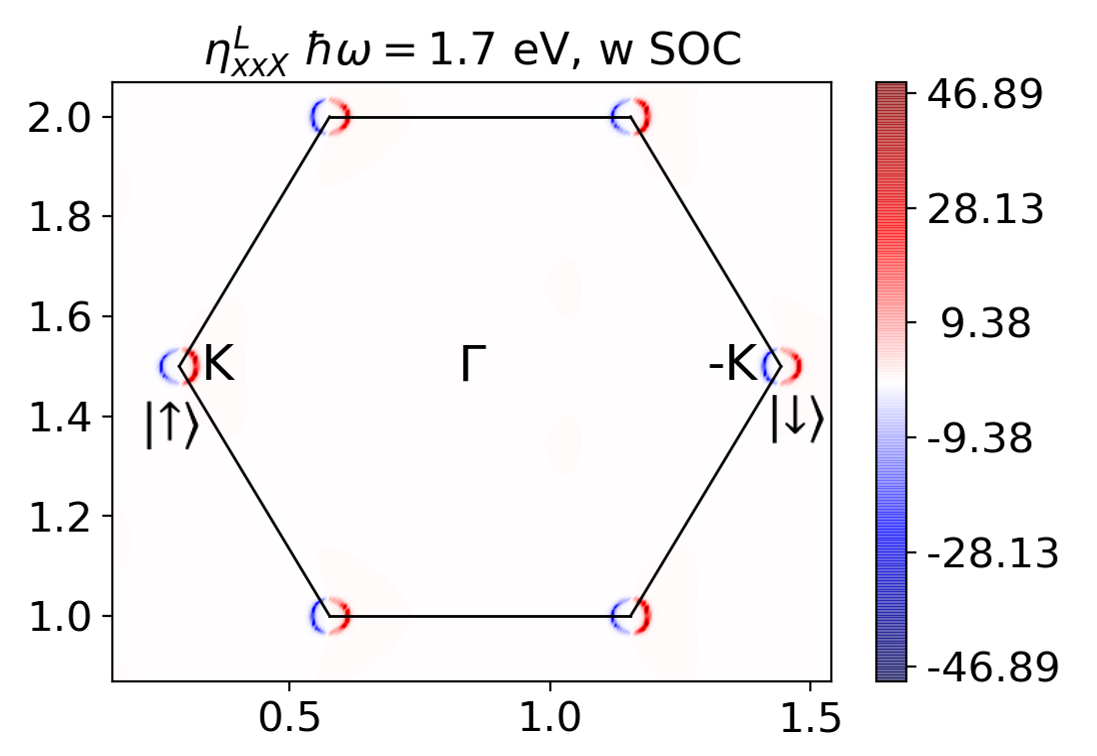}
\caption{
The spin current conductivity distributed 
in the reciprocal space. The K valley relate to 
the spin-up carrier excitation, while the -K 
valley relate to the spin-down carrier excitation at
photon E=1.7 eV.
}
\label{FigS4} 
\end{figure}

Fig.\ref{FigS4} shows the spin current conductivity
$\eta_{xxX}^L$ of monolayer MoS$_2$ distributed 
in the reciprocal space.
We plot the conductivity excited by the photon
at $E=1.7$ $eV$, which relates to only single spin
excitation at each valley. We can clear see that 
the spin-down and spin-up carrier has the opposite
sign, representing that the two spins travel in 
opposite direction. All the other TMDs is similar to 
that of MoS$_2$.
\subsection{Charge current under circular polarization}

\begin{figure}[!htbp]
\includegraphics[scale=0.22]{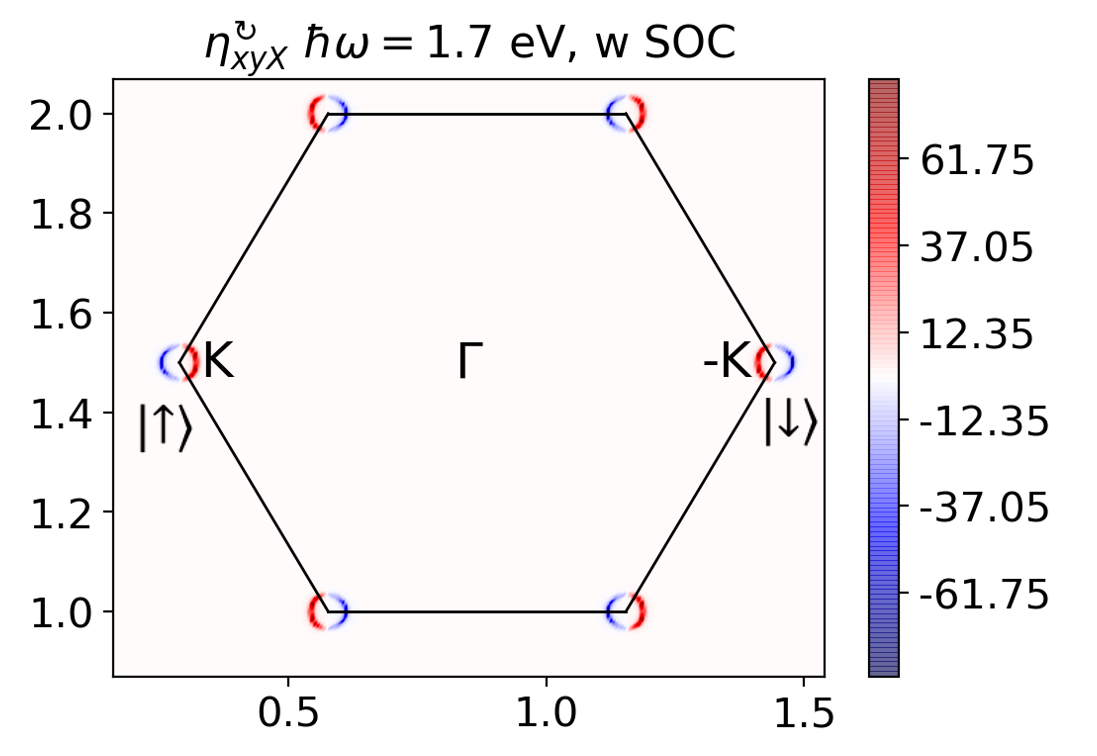}
\caption{
The charge current conductivity distributed 
in the reciprocal space. The K valley relate to 
the spin-up carrier excitation, and the -K 
valley relate to the spin-down carrier excitation at
photon E=1.7 eV.
}
\label{FigS5} 
\end{figure}

Fig.\ref{FigS5} shows the charge current conductivity
$\eta_{X}^{\circlearrowright}$ of monolayer MoS$_2$ distributed 
in the reciprocal space.
We also plot the conductivity excited by the photon
at $E=1.7$ $eV$. We note that 
the spin-down and spin-up carrier has the same sign, so the two spins travel in 
same direction.

\bibliography{supplement}